\input epsf.tex
\documentclass[12pt]{article}
\usepackage{graphicx}
\usepackage{psfig,picinpar,floatflt,amssymb,epsf}
\csname @addtoreset\endcsname{equation}{section}

\def\bseq{\begin{subequation}}  % = 1a 1b
\def\eseq{\end{subequation}}
\def\bsea{\begin{subeqnarray}}  % = 1.1a 1.1b
\def\esea{\end{subeqnarray}}
\newcommand{\beq}{\begin{equation}}
\newcommand{\eeq}{\end{equation}}
\newcommand{\bea}{\begin{eqnarray}}
\newcommand{\eea}{\end{eqnarray}}
\newcommand{\ena}{\end{eqnarray}}

\renewcommand{\a}{\alpha}
\renewcommand{\b}{\beta}

\renewcommand{\d}{\delta}
\newcommand{\pa}{\partial}
\newcommand{\g}{\gamma}
\newcommand{\G}{\Gamma}

\newcommand{\e}{\epsilon}

\newcommand{\phib}{\bar{\phi}}
\newcommand{\Phib}{\bar{\Phi}}
\newcommand{\adot}{\dot{\alpha}}
\newcommand{\bdot}{\dot{\beta}}

\newcommand{\thb}{\bar{\theta}}

\def\Mb{\kern 2pt\mathchoice
        {%displaystyle
         \vbox{\hrule width10pt height 0.4pt depth 0pt
         \kern 1.2pt\hbox{\kern -2pt$\displaystyle M$}}}
        {%textstyle
         \vbox{\hrule width10pt height 0.4pt depth 0pt
         \kern 1.2pt\hbox{\kern -2pt$\textstyle M$}}}
        {%scriptstyle \kern 0.5pt
\vbox{\hrule width6pt height 0.4pt depth 0pt
         \kern 1.0pt\hbox{\kern -2pt$\scriptstyle M$}}}
        {%scriptscriptstyle \kern 0.5pt
         \vbox{\hrule width5pt height 0.4pt depth 0pt
         \kern 0.8pt\hbox{\kern -2pt$\scriptscriptstyle M$}}}}

\def\Sb{\kern 2pt\mathchoice
        {%displaystyle
         \vbox{\hrule width6pt height 0.4pt depth 0pt
         \kern 1.2pt\hbox{\kern -2pt$\displaystyle S$}}}
        {%textstyle
         \vbox{\hrule width6pt height 0.4pt depth 0pt
         \kern 1.2pt\hbox{\kern -2pt$\textstyle S$}}}
        {%scriptstyle
         \vbox{\hrule width3.5pt height 0.4pt depth 0pt
         \kern 1.0pt\hbox{\kern -2pt$\scriptstyle S$}}}
        {%scriptscriptstyle
         \vbox{\hrule width3pt height 0.4pt depth 0pt
         \kern 0.8pt\hbox{\kern -2pt$\scriptscriptstyle S$}}}}

\def\Rb{\kern 2pt\mathchoice
        {%displaystyle
         \vbox{\hrule width5.5pt height 0.4pt depth 0pt
         \kern 1.2pt\hbox{\kern -2.5pt$\displaystyle R$}}}
        {%textstyle
         \vbox{\hrule width5.5pt height 0.4pt depth 0pt
         \kern 1.2pt\hbox{\kern -2.5pt$\textstyle R$}}}
        {%scriptstyle
         \vbox{\hrule width3.5pt height 0.4pt depth 0pt
         \kern 1.0pt\hbox{\kern -2.2pt$\scriptstyle R$}}}
        {%scriptscriptstyle
         \vbox{\hrule width3pt height 0.4pt depth 0pt
         \kern 0.8pt\hbox{\kern -2.2pt$\scriptscriptstyle R$}}}}

  \def\pp{{\mathchoice
      {%displaystyle
          \kern 1pt%
          \raise 1pt
          \vbox{\hrule width5pt height0.4pt depth0pt
            \kern -2pt
            \hbox{\kern 2.3pt
              \vrule width0.4pt height6pt depth0pt
              }
            \kern -2pt
            \hrule width5pt height0.4pt depth0pt}%
            \kern 1pt
       }
        {%textstyle
          \kern 1pt%
          \raise 1pt
          \vbox{\hrule width4.3pt height0.4pt depth0pt
            \kern -1.8pt
            \hbox{\kern 1.95pt
              \vrule width0.4pt height5.4pt depth0pt
              }
            \kern -1.8pt
            \hrule width4.3pt height0.4pt depth0pt}%
            \kern 1pt
        }
        {%scriptstyle
          \kern 0.5pt%
          \raise 1pt
          \vbox{\hrule width4.0pt height0.3pt depth0pt
            \kern -1.9pt  %[e]=0.15pt
            \hbox{\kern 1.85pt
              \vrule width0.3pt height5.7pt depth0pt
              }
            \kern -1.9pt
            \hrule width4.0pt height0.3pt depth0pt}%
            \kern 0.5pt
        }
        {%scriptscriptstyle
          \kern 0.5pt%
          \raise 1pt
          \vbox{\hrule width3.6pt height0.3pt depth0pt
            \kern -1.5pt
            \hbox{\kern 1.65pt
              \vrule width0.3pt height4.5pt depth0pt
              }
            \kern -1.5pt
            \hrule width3.6pt height0.3pt depth0pt}%
            \kern 0.5pt%}
        }
    }}

  \def\mm{{\mathchoice
               {%displaystyle
                 \kern 1pt
           \raise 1pt    \vbox{\hrule width5pt height0.4pt depth0pt
                  \kern 2pt
                  \hrule width5pt height0.4pt depth0pt}
                 \kern 1pt}
               {%textstyle
                \kern 1pt
           \raise 1pt \vbox{\hrule width4.3pt height0.4pt depth0pt
                  \kern 1.8pt
                  \hrule width4.3pt height0.4pt depth0pt}
                 \kern 1pt}
               {%scriptstyle
                \kern 0.5pt
           \raise 1pt
                \vbox{\hrule width4.0pt height0.3pt depth0pt
                  \kern 1.9pt
                  \hrule width4.0pt height0.3pt depth0pt}
                \kern 1pt}
               {%scriptscriptstyle
               \kern 0.5pt
         \raise 1pt  \vbox{\hrule width3.6pt height0.3pt depth0pt
                  \kern 1.5pt
                  \hrule width3.6pt height0.3pt depth0pt}
               \kern 0.5pt}
               }}

\def\pd{{\kern0.5pt
           + \kern-5.05pt \raise5.8pt\hbox{$\textstyle.$}\kern
0.5pt}}

\def\pmd{{\kern0.5pt
          \pm \kern-5.05pt
\raise6.3pt\hbox{$\textstyle.$}\kern1.5pt}}

\def\md{{\mathchoice
   {%displaystyle
      {{\kern 1pt - \kern-6.2pt \raise5pt\hbox{$\textstyle.$}\kern
1pt}}}
    {%textstyle
      {{\kern 1pt - \kern-6.2pt \raise5pt\hbox{$\textstyle.$}\kern
1pt}}}
    {%scriptstyle
      {\kern0.5pt - \kern-5.05pt
\raise3.4pt\hbox{$\textstyle.$}\kern0.5pt}}
    {%scriptscriptstyle
      {\kern0.5pt - \kern-5.05pt
\raise3.4pt\hbox{$\textstyle.$}\kern0.5pt}}}}

\begin{document}

\begin{titlepage}
{\hbox to\hsize{July  2003 \hfill
{Bicocca--FT--03--20}}}

\begin{center}
\vglue .06in
{\Large\bf Two-loop Renormalization for Nonanticommutative\\
$N=\frac12$ Supersymmetric WZ Model}
\\[.45in]
Marcus T. Grisaru\footnote{grisaru@physics.mcgill.ca}\\
{\it Physics Department, McGill University \\
Montreal, QC Canada H3A 2T8 }
\\
[.2in]
Silvia Penati\footnote{silvia.penati@mib.infn.it} ~and~
Alberto Romagnoni\footnote{alberto.romagnoni@mib.infn.it}\\
{\it Dipartimento di Fisica dell'Universit\`a degli studi di
Milano-Bicocca,\\
and INFN, Sezione di Milano, piazza della Scienza 3, I-20126 Milano,
Italy}\\[.8in]

{\bf ABSTRACT}\\[.0015in]
\end{center}
We study systematically, through   two loops,  the divergence structure of the
supersymmetric WZ model defined on the $N=\frac12$ nonanticommutative
superspace. By introducing a spurion field to represent the supersymmetry
breaking term $F^3$ we are able to perform our calculations  using conventional
supergraph techniques. Divergent terms proportional to $F$, $F^2$ and $F^3$
are produced (the first two are to be expected on general grounds)
 but no higher-point divergences are found. By adding
{\em ab initio} $F$ and $F^2$ terms to the original lagrangian  we
render the model renormalizable. We determine the renormalization constants
and beta functions through two loops, thus making it possible
 to study the renormalization group flow of the
nonanticommutation parameter.

${~~~}$ \newline
PACS: 03.70.+k, 11.15.-q, 11.10.-z, 11.30.Pb, 11.30.Rd  \\[.01in]
Keywords: Noncommutative geometry, $N=\frac{1}{2}$ Supersymmetry, Wess-Zumino model.

\end{titlepage}

\section{Introduction}

In the past few years, the properties of field theories defined over a
noncommutative (NC) space-time have been studied extensively, both at the
classical and at the quantum level (for a review and references see for instance
\cite{NC}). More recently, extensions of noncommutative geometry ideas to
superspace have been considered \cite{ferrara, KPT, OV, seiberg} and
investigations of field theories defined over such superspaces have been
initiated, both at the classical level \cite{seiberg}, at the quantum level
\cite{Rey, TY} and in connection with matrix models \cite{OV, park, Rey2}. Although the most general non(anti)commutative (N(A)C)
extensions involve both space-time $x^a$ and spinor $\theta^\a$ coordinates,
the simplest case is that of N=1 supersymmetric theories where the only
modification,  in a suitably defined {\it euclidean space}, of the ordinary
geometry involves the anticommutator $\{\theta^\a , \theta^\b \} = 2 C^{\a \b}$
with $C$ a nonzero constant, all the other (anti)commutators keeping their
usual values.  For this case Seiberg \cite{seiberg} has described extensions of
the usual field theories of chiral (scalar multiplet) and real scalar (vector
multiplet) superfields, and in refs. \cite{Rey, TY} some of the quantum
properties of the NC Wess-Zumino model have been investigated at low-loop
orders.

The effect of $\theta$-nonanticommutativity for the WZ model is very easy to
describe; the action, written in terms of ordinary component fields, is the
usual WZ component action augmented by a term proportional to $F^3$, cubic in
the auxiliary field (but without a corresponding $\bar{F}^3$). The $N=1$
supersymmetry is broken down to  $N=1/2$, and some of the remarkable quantum
properties of supersymmetric theories such as the standard nonrenormalization
theorems are no longer valid.  Other features such as the stability of the
vacuum energy and the existence of an antichiral ring, are unchanged.  This has
been demonstrated in the explicit examples worked out in refs. \cite{Rey,
TY} using superspace or component calculations.

In the present work, initiated after Seiberg's paper appeared, we cover some of
the same  ground but we are concerned primarily with the perturbative
renormalizability properties of the model. To two-loop order we show that,
although new divergences are generated, the model is renormalizable provided we
augment the NC WZ component action by  terms proportional to  $F$ and $F^2$.\footnote{
On general grounds, unless forbidden by some symmetry, such terms may be expected to
accompany higher powers such as $F^3$.} In
particular, the NC parameter $C_{\a \b}$ gets renormalized; by computing the
corresponding $\beta$-function one can follow its RG flow.

We have found it convenient to use a {\it spurion} field $U$
\cite{softbreaking} to generate the supersymmetry breaking term $F^3$ (and, subsequently,
$F^2$ and $F$); this allows us to use standard supergraph methods and ordinary
$D$-algebra techniques \cite{superspace} to perform all our calculations in
superspace.

Our paper is organized as follows: In the first Section we review the
$N=\frac12$ euclidean NAC superspace \cite{seiberg} and clarify its relation to
previous proposals \cite{KPT}. In the second Section, for the NAC WZ model, we
compute the divergent one--loop contributions and show that the model is
renormalizable if we add an $F^2$ term (together with a tadpole $F$)  to the
classical lagrangian. Two--loop divergent contributions with the insertion of
these new vertices are then computed in Section 3 where we show that the theory
is renormalizable at that order. In Section 4 we discuss the renormalization at
two loops and compute the beta functions for the couplings of the theory. The
last Section is then devoted to some conclusions. Two appendices are added
where we discuss in details the degree of divergence for the most general
one--loop and two--loop diagrams with an arbitrary number of insertions of the
spurion field $U$ (arbitrary power in the NAC parameter). There we show that,
at least up to two loops, only diagrams with a {\it single} insertion of $U$
can be divergent.

\section{$N=\frac12$ non(anti)commutative superspace}

It has been recently shown \cite{peter, OV, seiberg, seiberg2} that the IIB
superstring in the presence of a graviphoton background defines a superspace
geometry with nonanticommutative spinorial coordinates.
\footnote{Non-anticommutative
structures in field theory and gravity have been studied in
different contexts \cite{Moffat}.}

In \cite{KPT} the most general structure of non(anti)commutative superspaces
was discussed by studying the compatibility conditions between the presence of
nontrivial commutation relations for bosonic and/or fermionic variables and the
presence of supersymmetry. If we work in Minkowski signature, imposing the
extra condition for the algebra of the coordinates to be associative brings in
quite severe constraints which allow, as the only nontrivial
commutators, $[x, \theta]$,  $[x, \thb]$ and $[x,x]$. However, it was shown in
\cite{KPT} that  euclidean signature is less restrictive and a NAC superspace
with $\{ \theta, \theta \}$ different from zero can be defined consistently
with associativity.

Rigorously, a superspace with euclidean signature can be defined only when
extended susy is present because of the impossibility of assigning consistent
reality conditions for the pair of Weyl fermions $\theta_\a$, $\thb_{\adot}$
(for a detailed review on the subject see for instance \cite{vanproeyen}). This
is the reason why in \cite{KPT} $N=2$ euclidean NAC superspace was considered.
However, in the $N=1$ case one can still define a superspace with euclidean
signature by temporarly doubling the fermionic degrees of freedom
\footnote{S.P. acknowledges a discussion with N. Seiberg on this point.}. In
this context it is then clear that the description of $N=1$ euclidean
superspace is formally equivalent to euclidean $N=2$.

We briefly review the results of \cite{KPT}.
We describe $N=2$ euclidean superspace by coordinates
$(x^{\a \adot}, \theta^\a, \thb^\a, \theta^{\adot}, \thb^{\adot})$
subject to the complex conjugation conditions
\bea
&& (\theta^\a)^{\ast} = i \thb_\a  \qquad \quad ;
\qquad (\thb^\a)^{\ast} = -i \theta_\a
\nonumber \\
&& (\theta_\a)^{\ast} = -i \thb^\a  \qquad ;
\qquad (\thb_\a)^{\ast} = i \theta^\a
\label{hc}
\eea
and the same for dot variables. There are no h.c. relations between
$\theta^\a$ and $\theta^{\adot}$.

In \cite{KPT} we chose a nonchiral representation for the covariant spinor
derivatives and susy charges, but this choice brings us to a NAC algebra where
$\{ \theta, \theta \}$ different from zero necessarily implies nonvanishing
commutators between $\theta$'s and $x$'s.

Instead, if we use a chiral representation
(we consider only the left sector and use the conventions of \cite{superspace})
\bea
&& Q_\a = i ( \pa_\a - i \theta^{\adot} \pa_{\a \adot} )
\quad , \quad Q_{\adot} = i \pa_{\adot}
\nonumber \\
&& D_\a =  \pa_\a \qquad \qquad \qquad , \quad D_{\adot} = \pa_{\adot} + i
\theta^\a \pa_{\a \adot} \eea the susy transformations of the coordinates are
\beq
\d x^{\a \adot} = -i \e^\a \theta^{\adot} \quad , \quad
\d \theta^\a = \e^\a \quad , \quad \d \theta^{\adot} = \e^{\adot}
\label{susytransf}
\eeq
and the NAC algebra \beq \{
\theta^\a , \theta^\b \} = 2 C^{\a \b} \qquad {\rm the ~rest} = 0 \label{NC2}
\eeq with $C^{\a \b} = C^{\b \a}$ constant, is compatible with
(\ref{susytransf}) and  is associative. According to the general discussion in
\cite{KPT} it is easy to see that the algebra of derivatives gets modified as
\bea
&&\{D_\a , D_\b\}_\ast = 0 \qquad , \qquad  \{D_\a , D_{\adot} \}_\ast = i
\pa_{\a \adot}
\nonumber \\
&& \{D_{\adot} , D_{\bdot} \}_\ast = -2 C^{\a \b} \pa_{\a \adot} \pa_{\b \bdot}
\label{NC3}
\eea
while the algebra of the susy charges is not modified.
One might conclude that in this representation susy is not
broken. However, the modification of the anticommutation relations between
covariant derivatives makes it difficult to proceed and consistently define
(anti)chiral representations.

In \cite{seiberg} an alternative proposal was made which starts with
a different chiral representation for derivatives and charges
\bea
&& Q_{\adot} = i ( \pa_{\adot} - i \theta^{\a} \pa_{\a \adot} )
\quad , \quad Q_{\a} = i \pa_{\a}
\nonumber \\
&& D_{\adot} =  \pa_{\adot} \qquad \qquad \qquad , \quad
D_{\a} = \pa_{\a} + i \theta^{\adot} \pa_{\a \adot}
\label{DQ}
\eea
In principle the NAC algebra consistent with susy and associativity is
of the form
\bea
&& \{ \theta^\a , \theta^\b \} = 2 C^{\a \b} \qquad \{ \theta^{\adot} ,
\theta^\b \}=
\{ \theta^{\adot} , \theta^{\bdot} \} =0
\nonumber \\
&& [ x^{\a\adot} , \theta^\b ] = -2 i  C^{\a\b} \theta^{\adot}
\nonumber \\
&& [ x^{\a\adot} , x^{\b\bdot} ] = 2 \theta^{\adot} C^{\a\b} \theta^{\bdot}
\label{NC20}
\eea
but a suitable change of variable
\beq
y^{\a\adot} = x^{\a\adot} - i \theta^\a \theta^{\adot}
\label{y}
\eeq
avoids dealing with noncommuting $x$'s. Therefore the superspace described in
terms of $(y^{\a \adot}, \theta^\a, \thb^\a, \theta^{\adot}, \thb^{\adot})$
is dressed with a nonanticommutative geometry given by (\ref{NC2}).
In this case the algebra of the covariant spinor derivatives is not modified,
while
\bea
&&\{Q_\a , Q_\b\}_\ast = 0 \qquad , \qquad  \{Q_\a , Q_{\adot} \}_\ast =
i \pa_{\a \adot}
\nonumber \\
&& \{Q_{\adot} , Q_{\bdot} \}_\ast = 2 C^{\a \b} \pa_{\a \adot} \pa_{\b \bdot}
\label{NC30}
\eea
Therefore the supersymmetry is explicitly broken
\cite{seiberg} on the class of smooth functions defined on this superspace.
Supersymmetry seems to be broken in general to $N=\frac12$
\cite{seiberg, seiberg2, AIO}.
We note that the susy-breaking term is quadratic in the bosonic
derivatives, so
it does not spoil the previous statement about consistency of ({\ref{NC2}) with
supersymmetry invariance of the fundamental algebra of the coordinates.

Following Seiberg we realize the NAC geometry on the smooth superfunctions
defined on this superspace by introducing the nonanticommutative
(but associative) product
\bea
\phi \ast \psi &=& \phi e^{- \overleftarrow{\pa}_\a C^{\a \b}
\overrightarrow{\pa}_\b} \psi
\nonumber \\
&=& \phi \psi - \phi \overleftarrow{\pa}_\a C^{\a \b} \overrightarrow{\pa}_\b
\psi
+ \frac12 \phi \overleftarrow{\pa}_\a \overleftarrow{\pa}_\g C^{\a \b}
C^{\g \d}
\overrightarrow{\pa}_\d \overrightarrow{\pa}_\b \psi
\nonumber \\
&=& \phi \psi - \phi \overleftarrow{\pa}_\a C^{\a \b} \overrightarrow{\pa}_\b
\psi - \frac12 C^2 \pa^2\phi {\pa}^2 \psi
\label{star2}
\eea
where we have defined $C^2 =  C^{\a \b}C_{\a \b}$.
Since the covariant derivatives (\ref{DQ}) are still derivations for this
product, if we define (anti)chiral superfields as usual the classes of
(anti)chirals are still closed.

\section{The $N=\frac12$ WZ model: generalities}

On the non(anti)commutative superspace described in the previous Section we
define the WZ model as given by the ordinary cubic action where products of
superfields are generalized to the star product (\ref{star2}). We study the
model at the quantum level by performing its renormalization up to two loops.

We consider the classical action
\bea
&& S = \int d^8z \Phib \Phi - \frac{m}{2} \int d^6z \Phi^2
- \frac{\bar{m}}{2} \int d^6\bar{z} \Phib^2
\nonumber \\
&& - \frac{g}{3} \int d^6z \Phi \ast \Phi \ast \Phi - \frac{\bar{g}}{3} \int
d^6\bar{z} \Phib \ast \Phib \ast \Phib \label{action} \eea 
This action is generically complex since no h.c. relations are assumed for fields, masses and couplings.
Performing the expansion of the star product as in (\ref{star2}) and neglecting total superspace derivatives, the cubic interaction terms reduce to the usual WZ interactions augmented by the nonsupersymmetric component term $\frac{g}{6}\int d^4 x C^2 F^3 $. The action takes the form \cite{seiberg}
 \bea && S =
\int d^8z \Phib \Phi - \frac{m}{2} \int d^6z \Phi^2 - \frac{\bar{m}}{2} \int
d^6\bar{z} \Phib^2 - \frac{g}{3} \int d^6z \Phi^3  - \frac{\bar{g}}{3} \int
d^6\bar{z} \Phib^3
\nonumber \\
&& ~~~~+ \frac{g}{6} \int d^8z U (D^2 \Phi)^3 \label{action2} \eea where we
have introduced the external, constant spurion superfield $U = \theta^2 \thb^2
C^2$ in order to deal with a well-defined superspace expression for the extra
term proportional to the NC parameter. We note that an equivalent description
can be given where the extra term is expressed as $\int d^2
\theta \Phi (D^2 \Phi)^2$. However, the
integrand is not chiral and in principle it is not clear why it should be
inserted as an F-term in the action. Our choice allows us to use all the
standard tools and techniques of superspace perturbation theory.

 The action in
components reads ($\Phi| = \phi$, $D_\a \Phi| = \psi_\a$, $D^2 \Phi| = F$ and
analogously for the antichiral components) \bea &&S ~=~ \int d^4x \Big[ \phi
\Box \phib + F \bar{F}
-GF -\bar{G}\bar{F} + \frac{g}{6} C^2 F^3  \nonumber\\
&&~~~~~~~~~~~~~~~ + \psi^\a i \pa_\a^{\adot} \bar{\psi}_{\adot} - \frac{m}{2}
\psi^\a \psi_\a - \frac{\bar{m}}{2} \bar{\psi}^{\adot} \bar{\psi}_{\adot}
- g \phi \psi^\a \psi_\a - \bar{g} \phib \bar{\psi}^{\adot} \bar{\psi}_{\adot}
\Big]
\label{components}
\eea
where we have defined
\bea \label{GGbar}
&&  G = m \phi + g \phi^2 \nonumber \\
&& \bar{G}= \bar{m} \phib + \bar{g} \phib^2 \eea The auxilary fields $F$ and
$\bar{F}$ satisfy the algebraic equations of motion (EOM) \beq F = \bar{G}
\qquad , \qquad \bar{F} = G - \frac{g}{2} C^2 F^2 = G- \frac{g}{2} C^2
\bar{G}^2 \label{EOM} \eeq

We perform quantum--background splitting by setting $\Phi \rightarrow \Phi +
\Phi_q$ and integrating out the quantum fluctuations $\Phi_q$. The expansion
produces the ordinary quadratic and cubic vertices in $\Phi$ and $\Phib$ plus
two new extra vertices from the $U$ term. They are drawn in Fig. 1.

\vskip 18pt
\noindent
%---------- FIGURE TOP ------------
\begin{minipage}{\textwidth}
\begin{center}
\includegraphics[width=0.50\textwidth]{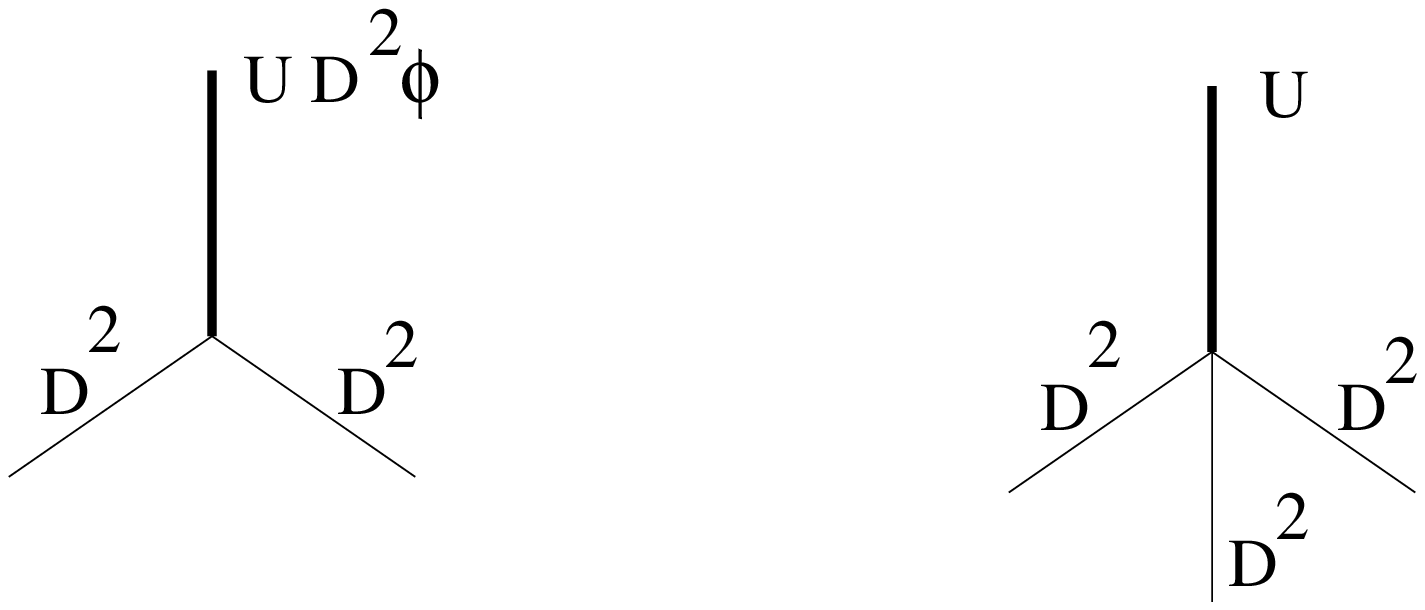}
\end{center}
\begin{center}
{\small{Figure 1: New vertices proportional to the external $U$ superfield}}
\end{center}
\end{minipage}
%---------- FIGURE END ------------

\vskip 20pt
The propagators are \cite{superspace}
\bea
&& \langle \Phi \Phib \rangle = \frac{1}{ p^2 + m \bar{m}}
\d^{(4)}(\theta - \theta')
\nonumber\\
&& \langle \Phi \Phi \rangle = - \frac{\bar{m} D^2}{ p^2(p^2 + m \bar{m})}
\d^{(4)}(\theta - \theta')
\nonumber\\
&& \langle \Phib \Phib \rangle = - \frac{m\bar{D}^2}{ p^2(p^2 + m\bar{m})}
\d^{(4)}(\theta - \theta') \eea Moreover, for each chiral (antichiral) field
there is an extra $\bar{D}^2$ ($D^2$) derivative on each line leaving a vertex
except for one of the lines at a (anti)chiral vertex.

At a given loop order we draw all supergraph configurations
with the corresponding chiral and antichiral derivatives from the
vertices and the propagators. Then we perform $D$--algebra to reduce
the supergraph to an ordinary momentum diagram.

We use BPHZ renormalization techniques. Thus, we start with the classical
action written in terms of renormalized quantities and order by order  perform
the subtraction of subdivergences directly on the diagrams. This procedure
takes  into account automatically the effect of insertion of lower order
counterterms.

We work in dimensional regularization ($n = 4 -2\e$) and minimal subtraction
scheme. It is convenient to regularize divergent integrals in the so-called
G--scheme \beq \int d^4k f(k) \rightarrow G(\e) \int d^n k f(k) \eeq where
$G(\e) = (4\pi)^{-\e} \G(1 -\e)$. A practical rule to deal with $4\pi$ factors
is to neglect them along the calculations and insert a $(4\pi)^2$ for each
momentum loop in the final result.

\section{One-loop divergences}

At one loop we have the ordinary self-energy $\Phi\Phib$ diagram which gives a
wave function renormalization. The divergent contribution is \beq A_{0}
~\rightarrow~ \frac{2}{\epsilon} g\bar{g} \int d^8z \Phi \bar{\Phi} \eeq New
divergent diagrams can appear which contain the $U$-vertices. In Appendix A we
study the most general one--loop diagram with a given number of $\Phi$, $\Phib$
external legs and an arbitrary number of $U$ vertices. We prove that at one
loop  diagrams with more than one insertion of $U$ vertices are convergent.
Moreover, with one $U$ insertion the only divergent topologies are the ones
given in Fig. 2. \vskip 18pt \noindent
%---------- FIGURE TOP ------------
\begin{minipage}{\textwidth}
\begin{center}
\includegraphics[width=0.60\textwidth]{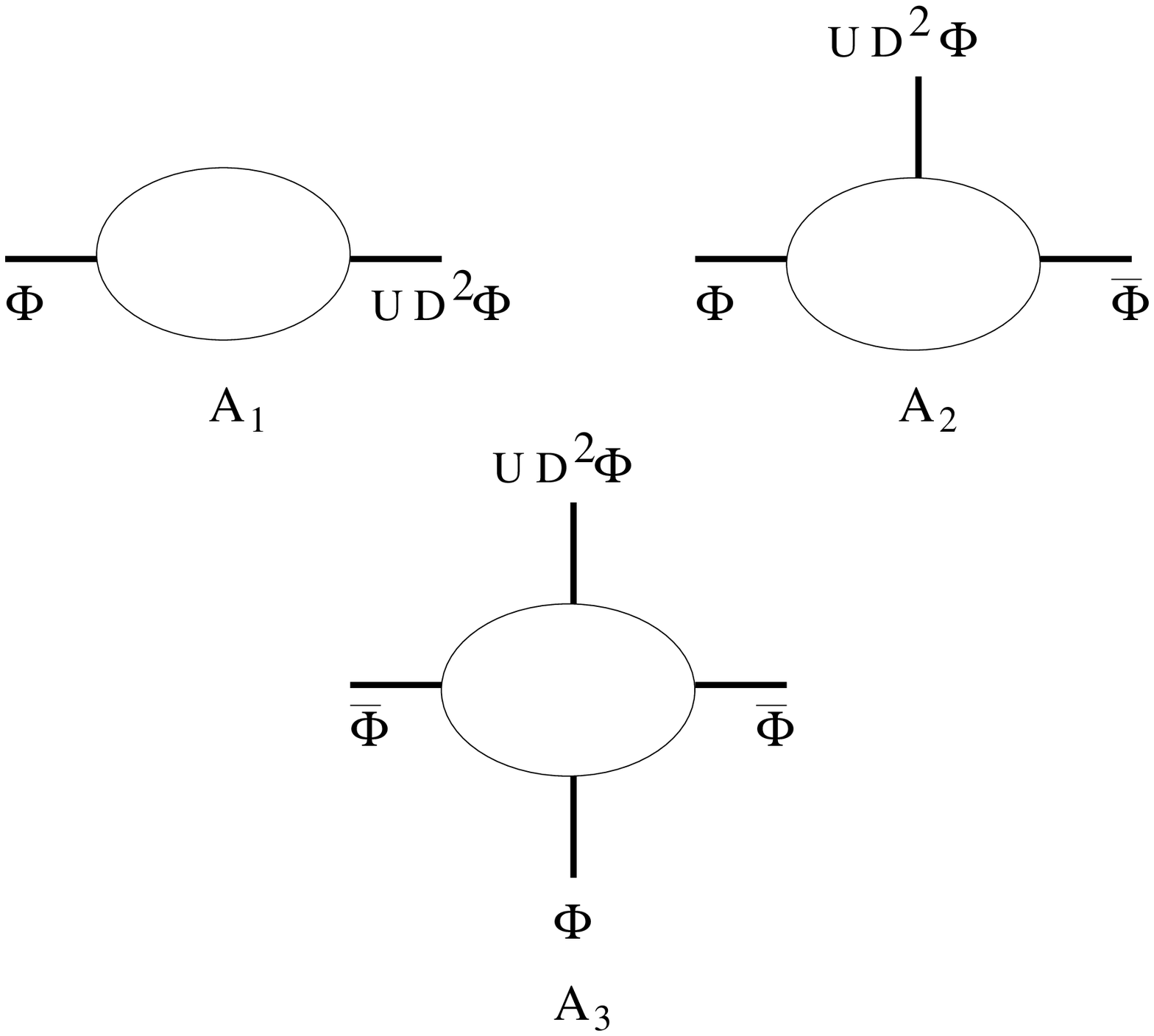}
\end{center}
\begin{center}
{\small{Figure 2:  One-loop divergent diagrams with one insertion of
the $U (D^2 \Phi)^3$-vertex}}
\end{center}
\end{minipage}
%---------- FIGURE END ------------

\vskip 20pt

Performing $D$--algebra and keeping only divergent terms all the
diagrams give rise to the self-energy momentum integral
\beq
\int d^4k \frac{1}{(k^2+ m\bar{m})[(p-k)^2+m\bar{m}]} \rightarrow
\frac{1}{\e}
\eeq
Computing the combinatorial factors the divergent contributions are
\bea
&& A_1 ~\rightarrow~  - \frac{1}{\e}
g^2 \bar{m}^2 \int d^8z U (D^2 \Phi)^2 ~=~
 - \frac{1}{\e} g^2 \bar{m}^2 C^2 \int d^4x ~F^2
\nonumber \\
&& A_2 ~\rightarrow~
- \frac{4}{\e} g^2 \bar{g} \bar{m} \int d^8z U (D^2 \Phi)^2 \Phib ~=~
 - \frac{4}{\e} g^2 \bar{g} \bar{m} C^2 \int d^4x F^2 \phib
\nonumber \\
&& A_3 ~\rightarrow~
- \frac{4}{\e} g^2 \bar{g}^2 \int d^8z U (D^2 \Phi)^2 \Phib^2 ~=~
 - \frac{4}{\e} g^2 \bar{g}^2 C^2 \int d^4x F^2 \phib^2
\label{counterterms} \eea In components their sum can be expressed as
\beq -
\frac{1}{\e} g^2 C^2 \int d^4x ~\Big[ \bar{m}^2~F^2 + 4 \bar{g} F^2 \bar{G}
\Big] \eeq
We note that, using the classical EOM (\ref{EOM}) the $\bar{G}$ in the second
term can be replaced by $F$
so that, in superfield form, the divergent contribution takes
the form \beq - \frac{1}{\e} g^2 \int d^8z \Big[ \bar{m}^2 U (D^2 \Phi)^2 + 4
\bar{g} U (D^2 \Phi)^3 \Big] \label{1loop} \eeq

The use of the classical EOM can be justified in the following manner: We have
started with the classical vertex proportional to $F^3$ and have produced, at
the one-loop level, a divergence which requires a counterterm proportional to
$F^2 \bar{G}$. This counterterm, to be added to the classical lagrangian, is to
be used now to cancel the one-loop divergence, but also, at a higher-loop
level, to remove subdivergences (which, following BPHZ, are removed by hand).
However we can show that inserting such a vertex into a diagram is completely
equivalent to inserting the vertex $F^3$. In fact,  let us consider the effect
of one field factor $F$ from such an insertion as compared to the factor
$\bar{G}=\bar{m} \phib + \bar{g} \phib^2$ or, which is more convenient,
$F-\bar{m} \phib$ as compared to $\bar{g} \phib^2$. We shall need the following
component propagators:
\bea
< F \bar{F}> &=& -\frac{\Box}{\Box - m \bar{m}} \nonumber\\
<\bar{\phi} \bar{F}> &=&  -\frac{m}{\Box - m \bar{m}} \nonumber\\
<{\phi} {F}> &=& - \frac{\bar{m}}{\Box - m \bar{m}} \nonumber\\
<\phi \bar{\phi}> &=&- \frac{1}{\Box - m \bar{m}} \eea In the Wick expansion,
the operators in $F-\bar{m} \phib$ can be contracted either
 with $\phi$
in the  cubic vertices $- {g}{\phi^2} {F}$ or
$- {g} \phi {\psi}^{\a} {\psi}_{\a}$ but given
the form of the propagators
the result is zero; or with the $\bar{F}$ in the cubic vertex
$-{\bar{g}}\bar{\phi}^2 \bar{F}$ and, given the form of the propagators,
the result is $1 \cdot \bar{g} \bar{\phi}^2$ thus establishing
 that   $F-\bar{m} \phib$ is completely equivalent to
  $\bar{g} \bar{\phi}^2$. Higher powers of these operators can be
  treated in the same manner (except for some combinatorial factors -- see
  Section 4) thus showing the equivalence of the two forms of
  counterterms when inserted into diagrams.

  Finally, in comparing {\it bare} lagrangians, i.e. the renormalized
  lagrangians plus counterterms, the equivalence of the two sets of
  counterterms follows just by using the corresponding equations of motion,
  in particular the equation for $\bar{F}$, that follow from these
  lagrangians.\footnote{We emphasize that the use of the classical EOM is valid
  only in the limited sense employed here. In general, the full {\it quantum}
  equations must be used when dealing with the full effective action.}

\vskip 20pt

\subsection{One--loop diagrams with the new term $U(D^2\Phi)^2$}

As shown above (see eq. (\ref{1loop})), at one loop a
divergent term proportional to $F^2$ appears which is not present in the
classical action. This implies that the theory described by (\ref{action}) is
not renormalizable. We  consider therefore a modified action with the addition
of $F$ and $F^2$ terms, written in superspace with the help of the spurion
field $U$ (although the $F$ term could also be written as a perfectly good
chiral integral of $\Phi$)
\bea && S = \int d^8z \Phib \Phi - \frac{m}{2} \int d^6z
\Phi^2 - \frac{\bar{m}}{2} \int d^6\bar{z} \Phib^2 - \frac{g}{3} \int d^6z
\Phi^3  - \frac{\bar{g}}{3} \int d^6\bar{z} \Phib^3
\nonumber \\
&& ~~~~+ \frac{g}{6} \int d^8z U (D^2 \Phi)^3 + k_1 \bar{m}^4
\int d^8z U D^2 \Phi + k_2 \bar{m}^2 \int d^8z U (D^2 \Phi)^2
\eea where we have introduced new dimensionless coupling constants $k_1$ and $
k_2$. This implies the insertion of mass powers. The choice to use $\bar{m}$ rather than $m$ is at this point completly arbitrary but it allows to simplify calculations. The presence of the  linear $F$ term  is required since linear
(divergent) contributions are necessarily produced at the quantum level once a
quadratic vertex is present. In principle it (or the $F^2$ term) could be absorbed
 by suitable field redefinitions but at the expense of producing other terms; we
 prefer to keep them explicitly in the classical action.

From the quadratic vertex $U(D^2\Phi)^2$ new topologies of diagrams emerge which
give rise to divergent contributions. By using a general procedure like the one
described in Appendix A we select the new divergent diagrams as given in
Fig. 3. Again, diagrams with more than one $U$-insertion are convergent. \vskip
18pt \noindent
%---------- FIGURE TOP ------------
\begin{minipage}{\textwidth}
\begin{center}
\includegraphics[width=0.65\textwidth]{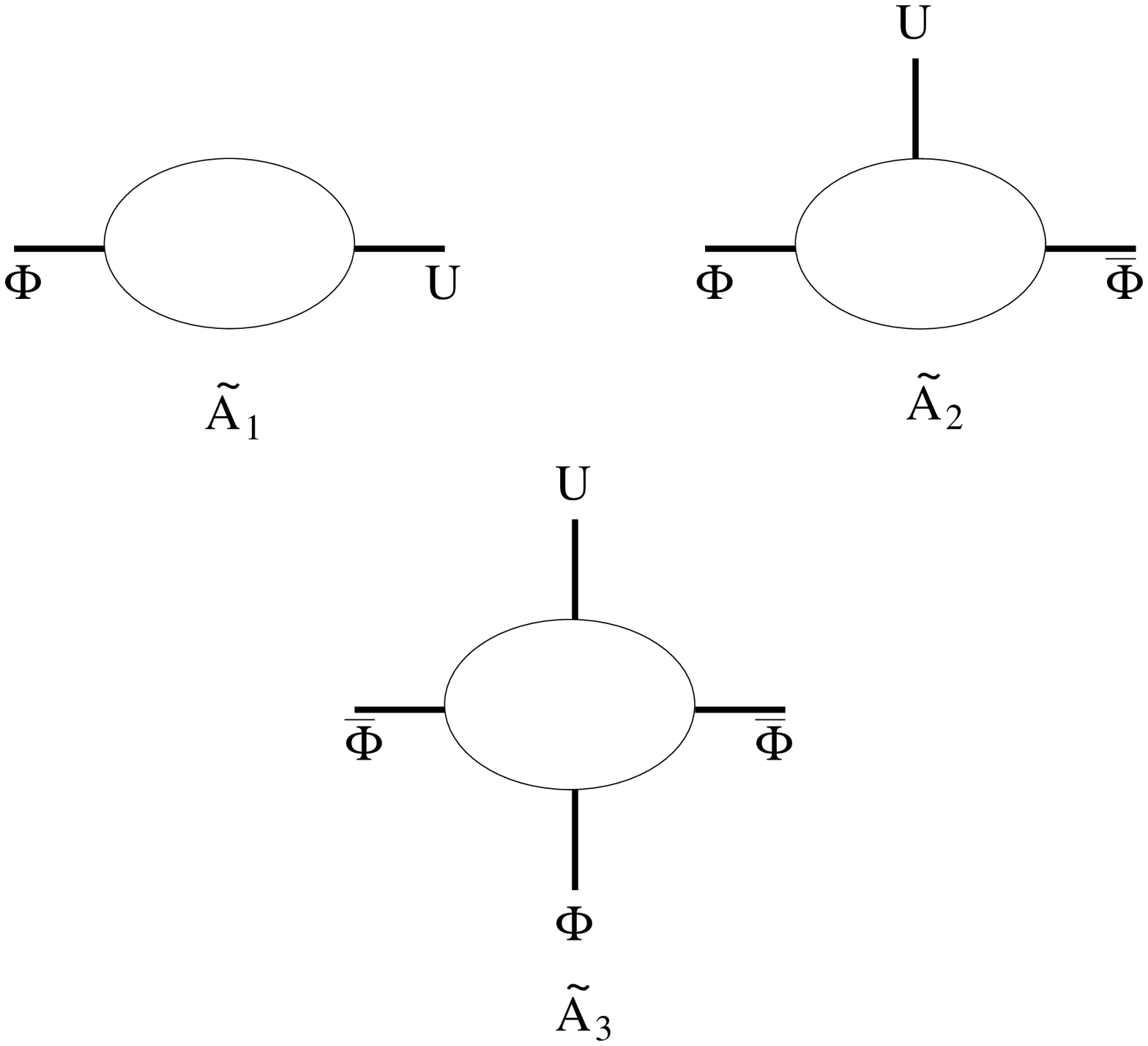}
\end{center}
\begin{center}
{\small{Figure 3: One-loop divergent diagrams with one insertion of
the $U (D^2 \Phi)^2$-vertex}}
\end{center}
\end{minipage}
%---------- FIGURE END ------------
\vskip 20pt

Computing the combinatorial factors, the  contributions are \bea &&
\widetilde{A}_1 ~\rightarrow~ - \frac{2}{\e} ~k_2~ g \bar{m}^4 \int d^8z U (D^2
\Phi) ~=~ - \frac{2}{\e} ~k_2~g \bar{m}^4 C^2 \int d^4x ~F
\nonumber \\
&& \widetilde{A}_2 ~\rightarrow~
- \frac{8}{\e}~k_2~ g \bar{g} \bar{m}^3 \int d^8z U (D^2\Phi) \bar{\Phi} ~=~
- \frac{8}{\e}~k_2~ g \bar{g} \bar{m}^3 C^2 \int d^4x F \phib
\nonumber \\
&& \widetilde{A}_3 ~\rightarrow~
- \frac{8}{\e}~k_2~ g \bar{g}^2 \bar{m}^2 \int d^8z U (D^2\Phi) \Phib^2 ~=~
- \frac{8}{\e}~ k_2~g \bar{g}^2 \bar{m}^2 C^2 \int d^4x F \phib^2
\eea
and they sum up to the following divergent expression
\beq
\widetilde{A}_1 + \widetilde{A}_2 + \widetilde{A}_3 ~=~
- \frac{2}{\e} k_2 g \bar{m}^2 C^2 \int d^4x \Big[ \bar{m}^2 F ~+~
4 \bar{g} F \bar{G} \Big]
\eeq
Using the classical EOM (\ref{EOM}) for the $F$-field the second term
is an $F^2$ contribution.

Therefore, summing everything and reinserting $(4\pi)$ factors,
the total one--loop divergence in superspace language is
\beq
- \frac{1}{\e} \frac{1}{(4\pi)^2}\int d^8z \Big[ 2 k_2 g \bar{m}^4 U(D^2\Phi)
+ \bar{m}^2 (g^2 + 8k_2 g\bar{g}) U (D^2 \Phi)^2 +
4 g^2 \bar{g} U (D^2 \Phi)^3 \Big]
\label{final1loop}
\eeq

\section{Two--loop divergences}

At two loops there is the ordinary self-energy $\Phi\Phib$ diagram which
induces a wave function renormalization.
New
divergent contributions can arise by $U$-insertions due to both quadratic
$U(D^2\Phi)^2$ and cubic $U(D^2\Phi)^3$ vertices. In Appendix B we give a detailed
analysis of potentially divergent diagrams and show that even at two loops
divergences can arise only when a single $U$ vertex (quadratic or cubic) is
present in the diagram. We list them by omitting diagrams which are convergent
after subtraction of subdivergences. Once $D$-algebra is performed,
nonvanishing divergent contributions always reduce to the following momentum
integrals (corresponding to the two configurations drawn in Fig. 4) \bea
I_{1}&=& \int \frac{d^{n}k d^{n}q}{[k^2 + m\bar{m}]~
[(p-k)^2 + m\bar{m}]~[q^2 + m\bar{m}]~[(k-q)^2 + m\bar{m}]} \nonumber \\
I_{2} &=& \int \frac{d^{n}k d^{n}q}{[k^2 + m\bar{m}]~[(p-k)^2 + m\bar{m}] ~[q^2
+ m\bar{m}]~[(p-q)^2 + m\bar{m}]} \label{int} \eea In dimensional
regularization, after subtraction of the self--energy subdivergences they give
\beq I_1 \rightarrow \quad - \frac{1}{2\e^2} + \frac{1}{2\e} \qquad, \qquad I_2
\rightarrow \quad - \frac{1}{\e^2} \eeq

%---------- FIGURE TOP ------------
\begin{minipage}{\textwidth}
\begin{center}
\includegraphics[width=0.70\textwidth]{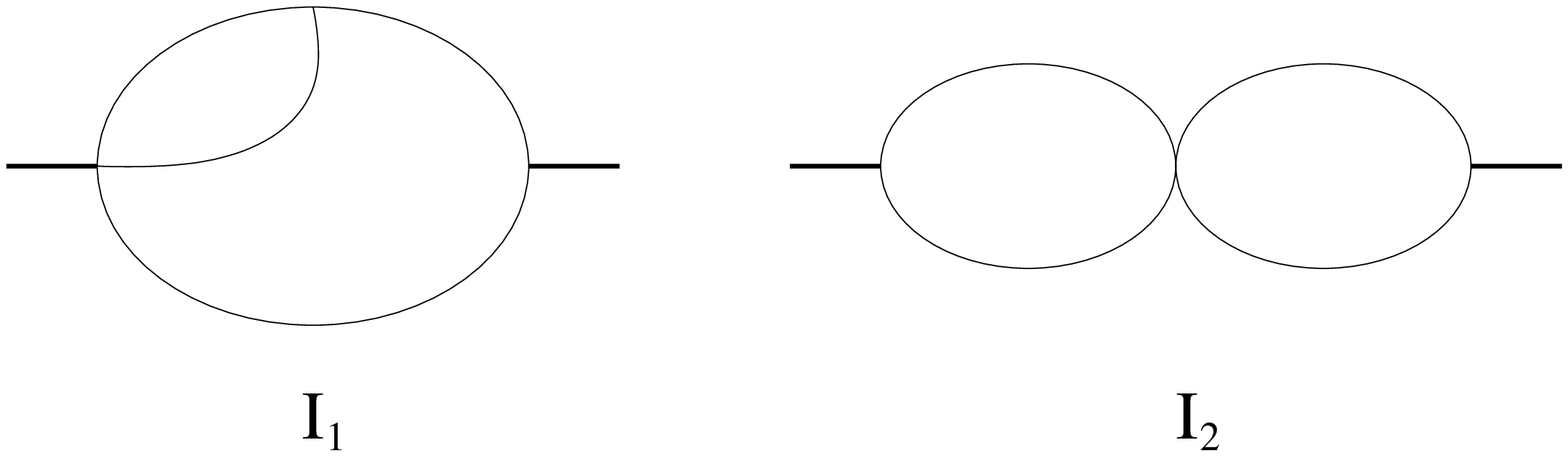}
\end{center}
\begin{center}
{\small{Figure 4: Divergent loop integrals}}
\end{center}
\end{minipage}
%---------- FIGURE END ------------

%%%%%%%%%%%%%%%%%%%%%%%%%%%%%%%%%%%%%%%

\vskip 18pt

The ordinary two--loop contribution to the self--energy $\Phi \Phib$ is then
\beq
- [8 I_{1}] g^2 \bar{g}^2 \int d^8z\Phi\bar{\Phi} = -
[- \frac{1}{\e^2}+\frac{1}{\e}] 4g^2 \bar{g}^2 \int d^8z\Phi\bar{\Phi}
\eeq
We now list all the other divergent contributions classifying them according to
the number of external fields. For each configuration we keep distinct
the diagrams with a cubic $U$ vertex from the ones with a quadratic one
(in the Figures they are indicated as tilde quantities).

\subsection{One-point functions}

At two loops we have the tadpoles drawn in Figs. 5 and  6
\vskip 18pt \noindent
%---------- FIGURE TOP ------------
\begin{minipage}{\textwidth}
\begin{center}
\includegraphics[width=0.25\textwidth]{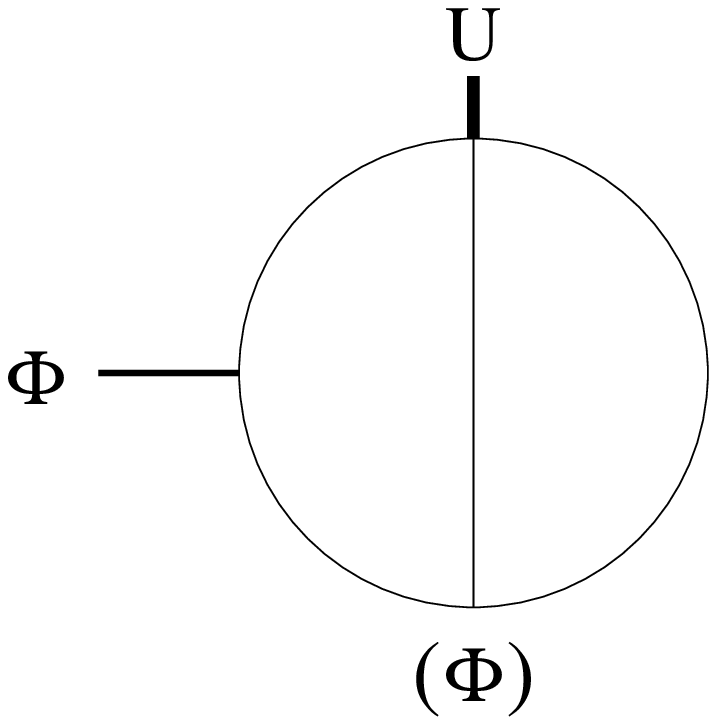}
\end{center}
\begin{center}
{\small{Figure 5:
Two--loop contribution $U(D^2\Phi) \quad \rightarrow \quad T_{1}$ }}
\end{center}
\end{minipage}
\vskip 10pt
\begin{minipage}{\textwidth}
\begin{center}
\includegraphics[width=0.70\textwidth]{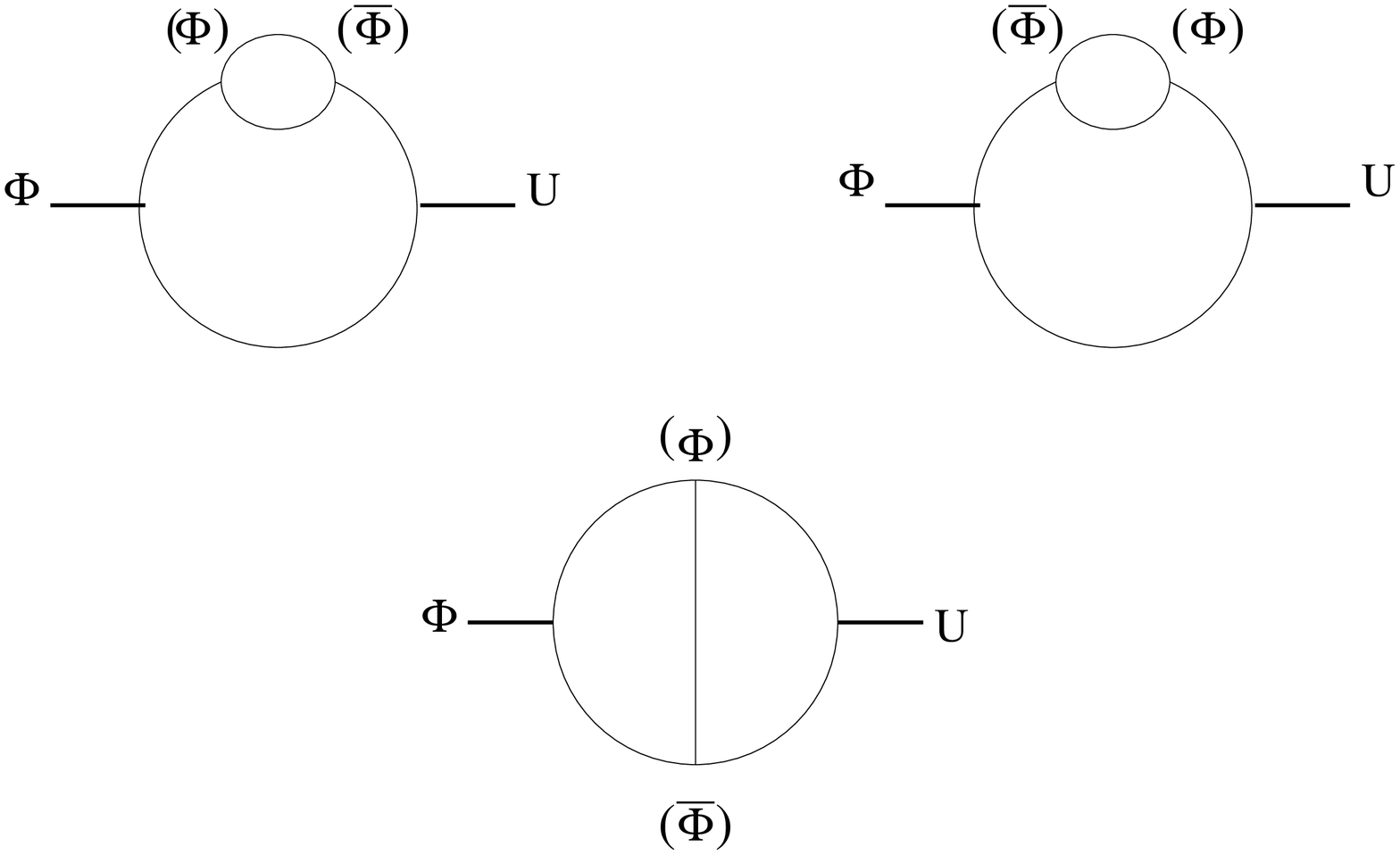}
\end{center}
\begin{center}
{\small{Figure 6: Two--loop contributions $U(D^2\Phi) \quad \rightarrow \quad \widetilde{T}_{1}$}}
\end{center}
\end{minipage}
%---------- FIGURE END ------------

\vskip 40pt

Computing the combinatorial factors they give
\bea
&&T_{1} ~:~  [2 I_{1}]g^3 \bar{m}^4 \int d^8z U (D^2\Phi) ~\rightarrow~
\left(- \frac{1}{\e^2}+\frac{1}{\e}\right) g^3 \bar{m}^4 C^2 \int d^4x ~F~
\nonumber \\
&&\widetilde{T}_{1} ~:~ [(8+8+16)I_{1}] k_2 g^2 \bar{g} \bar{m}^4
\int d^8z U (D^2\Phi)~\rightarrow~\left(- \frac{1}{\e^2}+\frac{1}{\e}\right)
16 k_2 g^2 \bar{g} \bar{m}^4 C^2 \int d^4x ~F\nonumber \\
\eea

%%%%%%%%%%%%%%%%%%%%%%%%%%%%%%%%%%%%%%%

\subsection{Two-point functions}

The divergent contributions to the two-point functions are given by
the graphs in Figs. 7--10.

\noindent
%---------- FIGURE TOP ------------
\begin{minipage}{\textwidth}
\begin{center}
\includegraphics[width=0.67\textwidth]{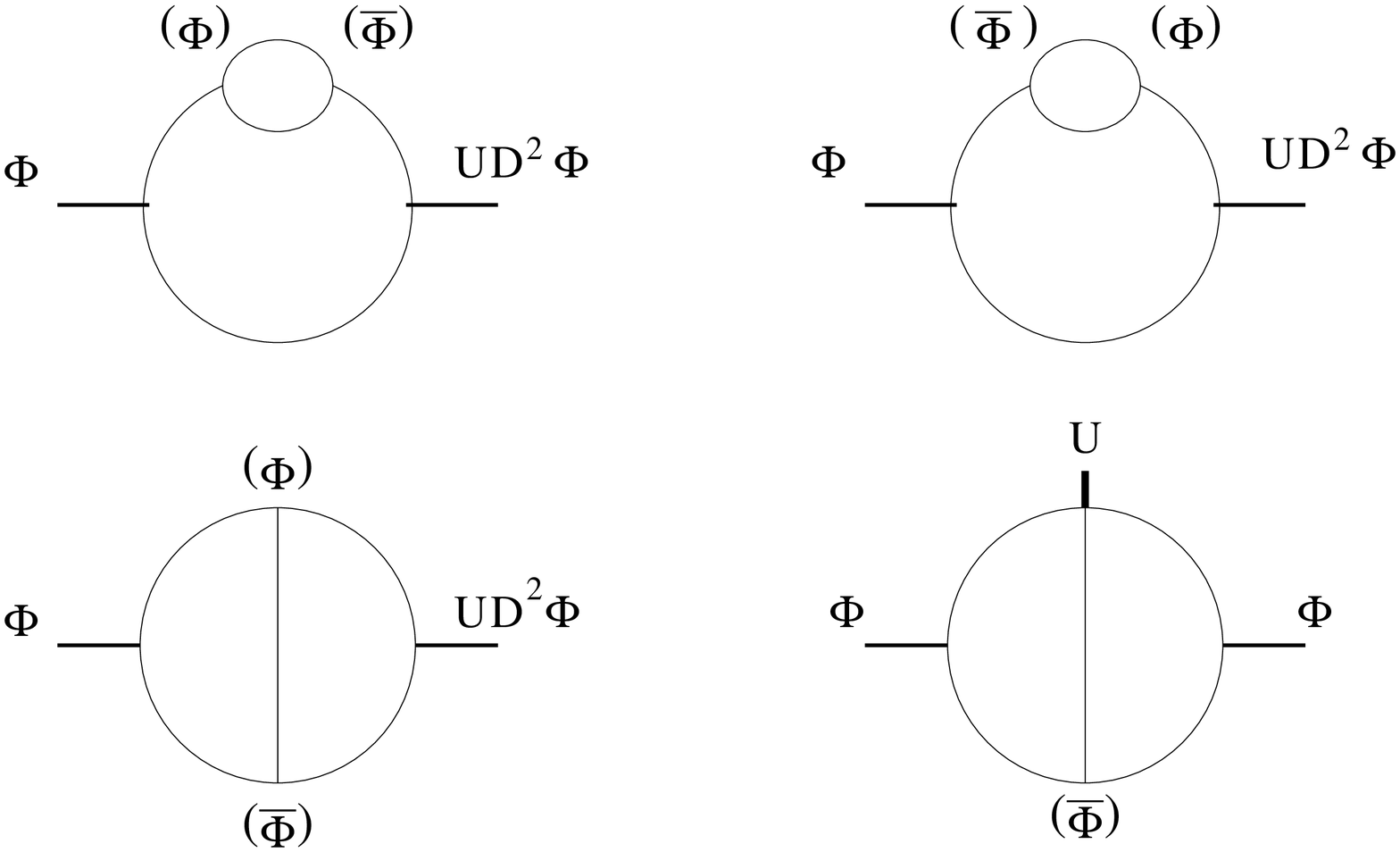}
\end{center}
\begin{center}
{\small{Figure 7: Two--loop contributions $U(D^2 \Phi)^2 \quad \rightarrow \quad B_{1}$}}
\end{center}
\end{minipage}
\vskip 10pt
\begin{minipage}{\textwidth}
\begin{center}
\includegraphics[width=0.30\textwidth]{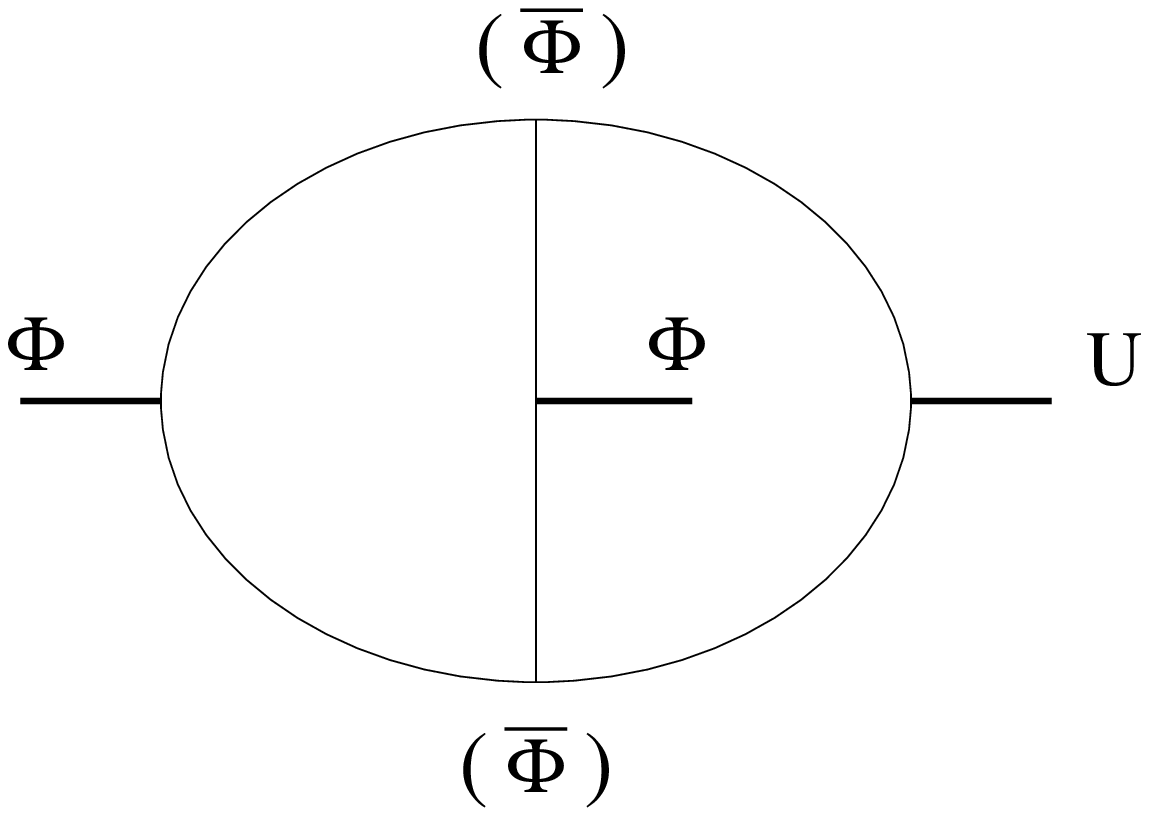}
\end{center}
\begin{center}
{\small{Figure 8: Two--loop contribution $U(D^2 \Phi)^2 \quad \rightarrow \quad \widetilde{B}_{1}$}}
\end{center}
\end{minipage}
\vskip 10pt
\begin{minipage}{\textwidth}
\begin{center}
\includegraphics[width=0.67\textwidth]{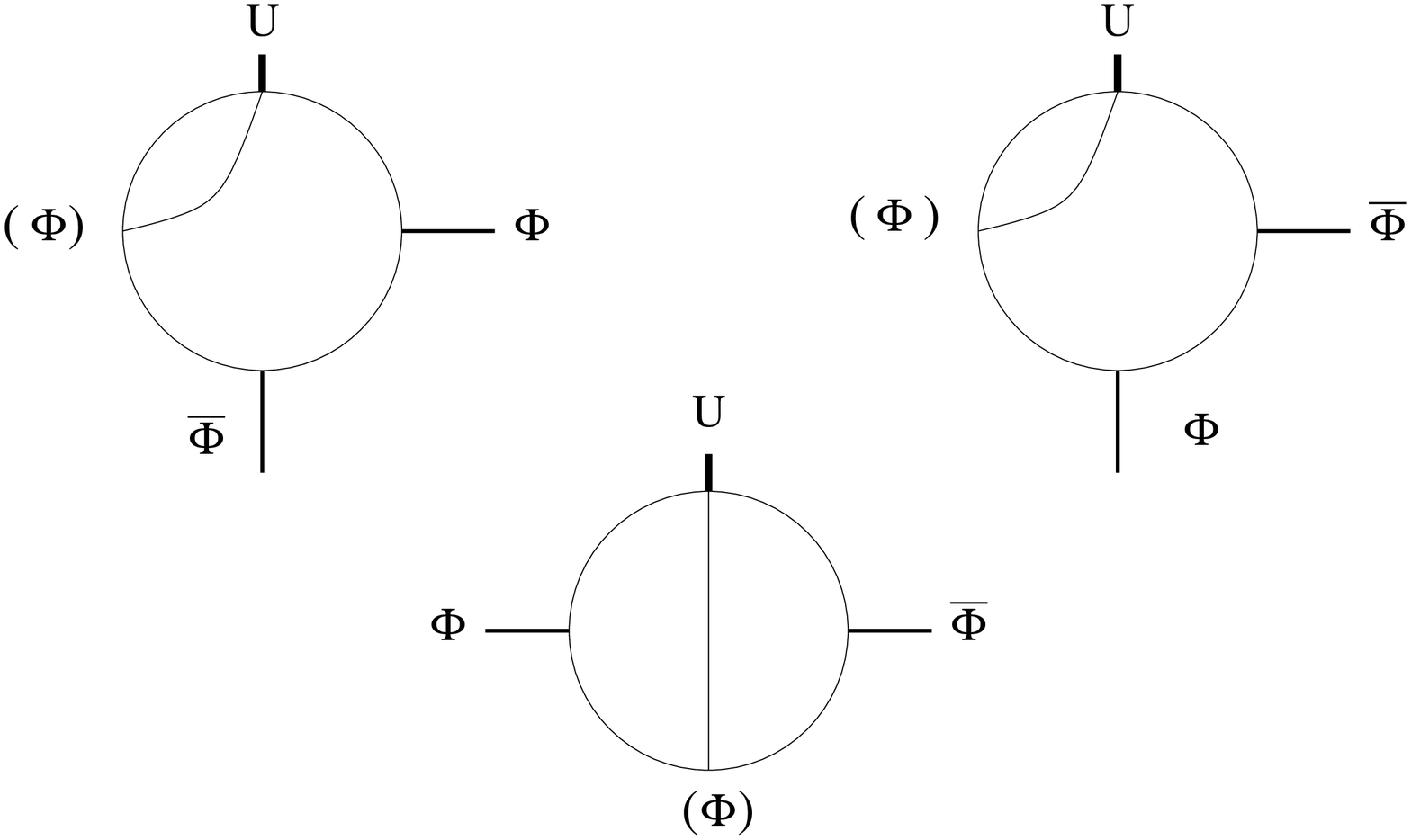}
\end{center}
\begin{center}
{\small{Figure 9: Two--loop contributions $U(D^2\Phi) \bar{\Phi} \quad \rightarrow \quad B_{2}$}}
\end{center}
\end{minipage}
\vskip 10pt
\begin{minipage}{\textwidth}
\begin{center}
\includegraphics[width=0.67\textwidth]{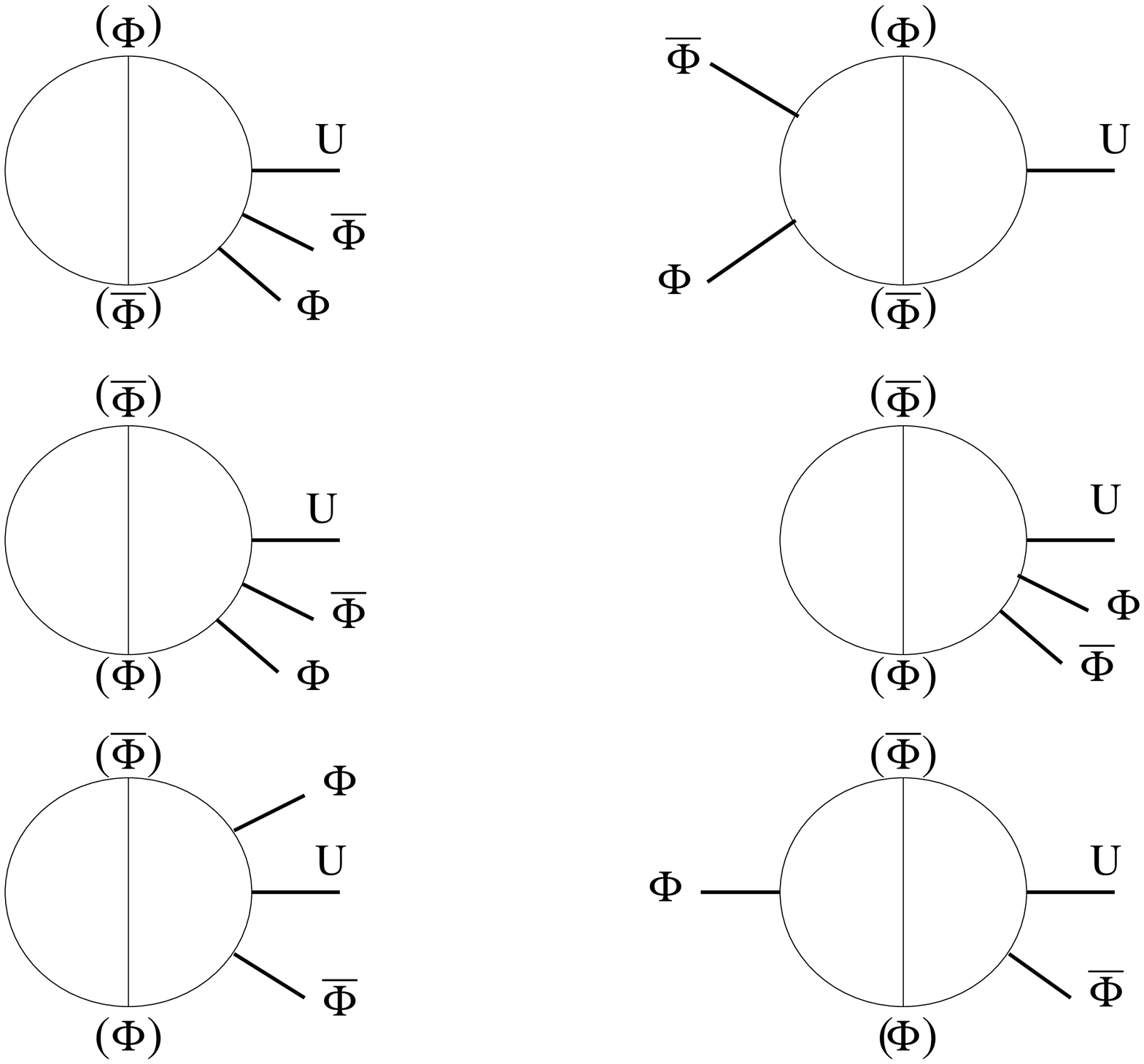}
\end{center}
\begin{center}
{\small{Figure 10: Two--loop contributions $U(D^2\Phi) \bar{\Phi} \quad
\rightarrow \quad \widetilde{B}_{2}$}}
\end{center}
\end{minipage}
%---------- FIGURE END ------------

\vskip 40pt

Computing the combinatorial factors, we have the following contributions
\bea
&&B_{1} ~:~  [(4+4+8)I_{1}+4I_{2}]g^3 \bar{g} \bar{m}^2 \int d^8z U(D^2
\Phi)^2 \nonumber \\
&& \hspace{235pt} \rightarrow~
\left(- \frac{3}{\e^2}+\frac{2}{\e}\right) 4g^3 \bar{g}
\bar{m}^2 C^2 \int d^4x ~F^2\nonumber \\
\nonumber \\
&&\widetilde{B}_{1} ~:~  [8I_{2}] k g^2 \bar{g}^2 \bar{m}^2 \int d^8z
U(D^2 \Phi)^2~\rightarrow~ - \frac{8}{\e^2} k_2 g^2 \bar{g}^2 \bar{m}^2
C^2
\int d^4x ~F^2\nonumber \\
\nonumber \\
&&B_{2} ~:~  [(4+4+8)I_{1}]g^3 \bar{g} \bar{m}^3 \int d^8z U (D^2\Phi)
\bar{\Phi} ~\rightarrow~  \left(- \frac{1}{\e^2}+\frac{1}{\e}\right)
8g^3 \bar{g} \bar{m}^3 C^2 \int d^4x ~F~\bar{\phi}\nonumber \\
&&\widetilde{B}_{2} ~:~  [(16+32+16+16+16+32)I_{1}]k g^2 \bar{g}^2 \bar{m}^3
\int d^8z U (D^2\Phi) \bar{\Phi} \nonumber\\
&& \hspace{235pt} \rightarrow~ \left(- \frac{1}{\e^2}+\frac{1}{\e}\right)
64k_2 g^2
\bar{g}^2 \bar{m}^3 C^2 \int d^4x ~F~\bar{\phi}\nonumber \\
\eea

%%%%%%%%%%%%%%%%%%%%%%%%%%%%%%%%%%%%%%%%%%%%

\subsection{Three-point function}

For the three-point functions the divergent diagrams are drawn in
Figs. 11--14.

\vskip 18pt
\noindent
%---------- FIGURE TOP ------------
\begin{minipage}{\textwidth}
\begin{center}
\includegraphics[width=0.33\textwidth]{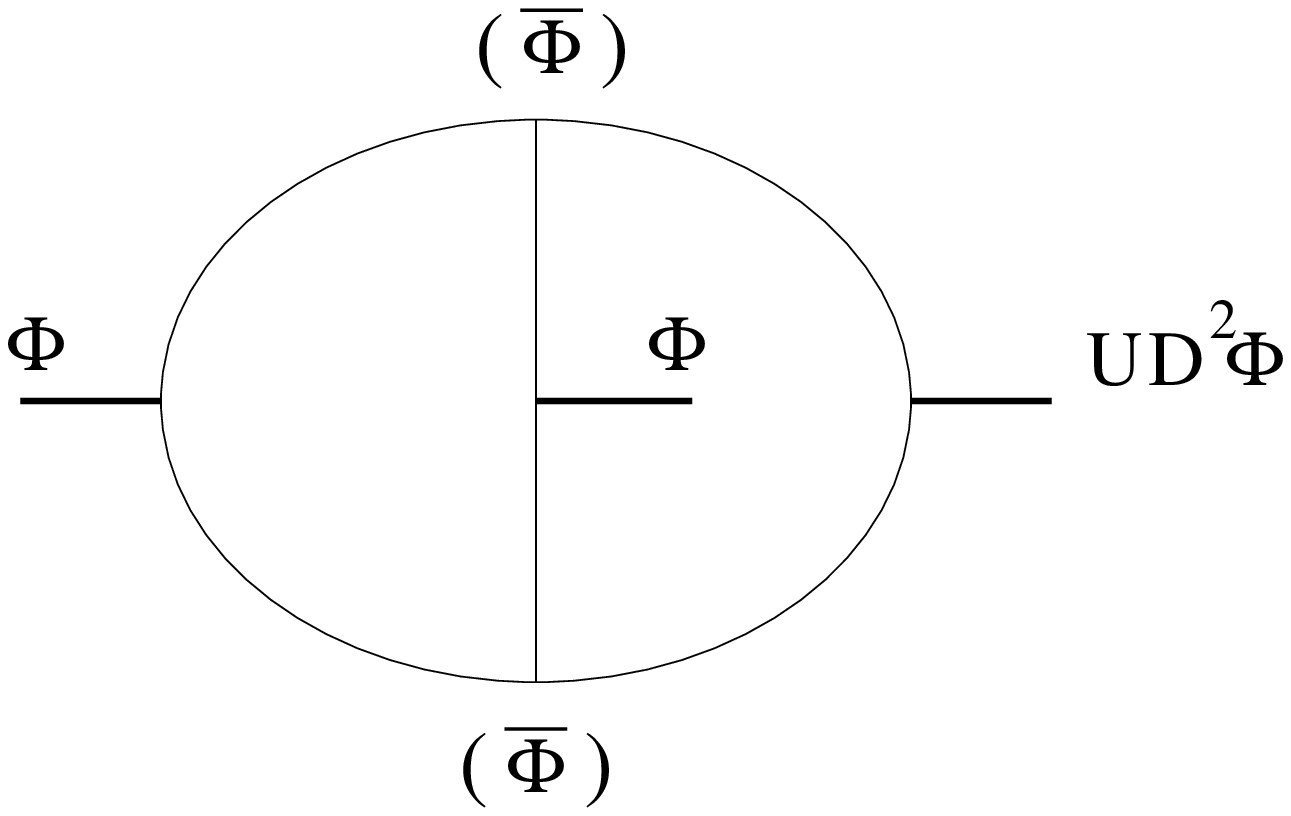}
\end{center}
\begin{center}
{\small{Figure 11: Two--loop contribution $U(D^2 \Phi)^3 \quad
\rightarrow \quad C_{1}$}}
\end{center}
\end{minipage}
\begin{minipage}{\textwidth}
\begin{center}
\includegraphics[width=0.62\textwidth]{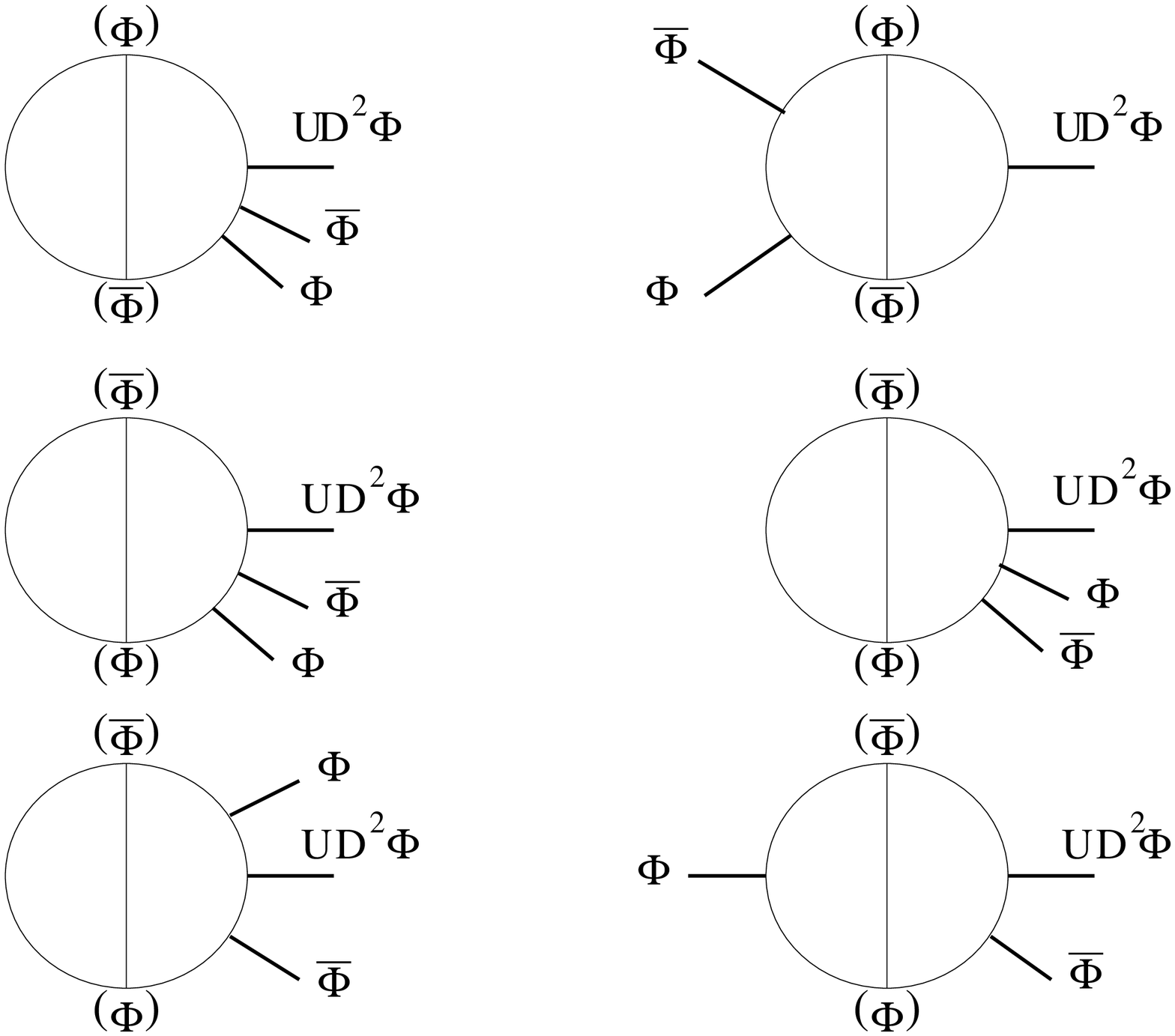}
\end{center}
\begin{center}
\includegraphics[width=0.26\textwidth]{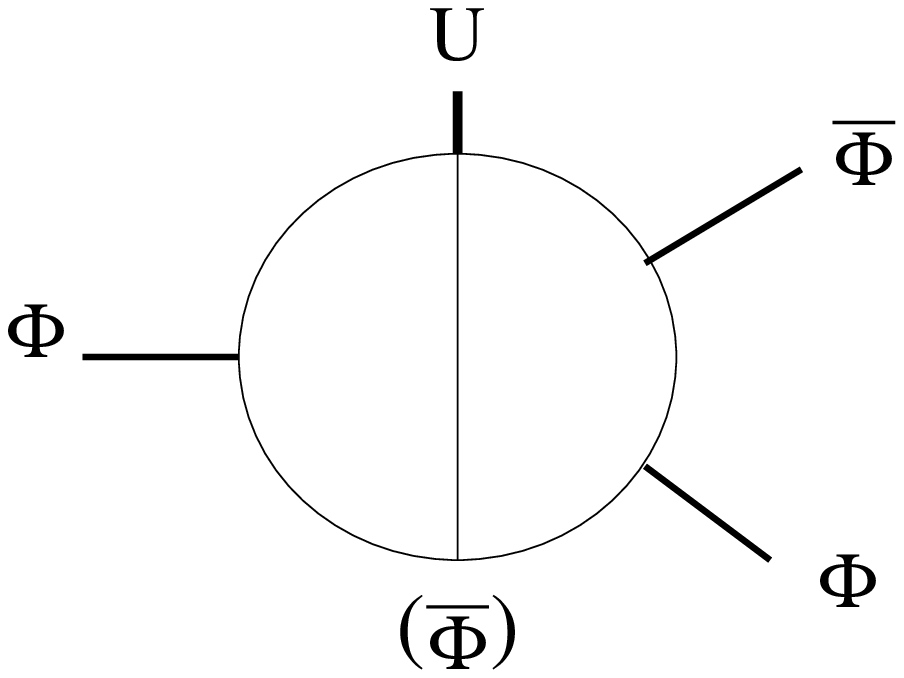}
\end{center}
\begin{center}
{\small{Figure 12: Two--loop contributions $U (D^2 \Phi)^2 \bar{\Phi} \quad \rightarrow \quad C_{2}$ }}
\end{center}
\end{minipage}
\begin{minipage}{\textwidth}
\begin{center}
\includegraphics[width=0.70\textwidth]{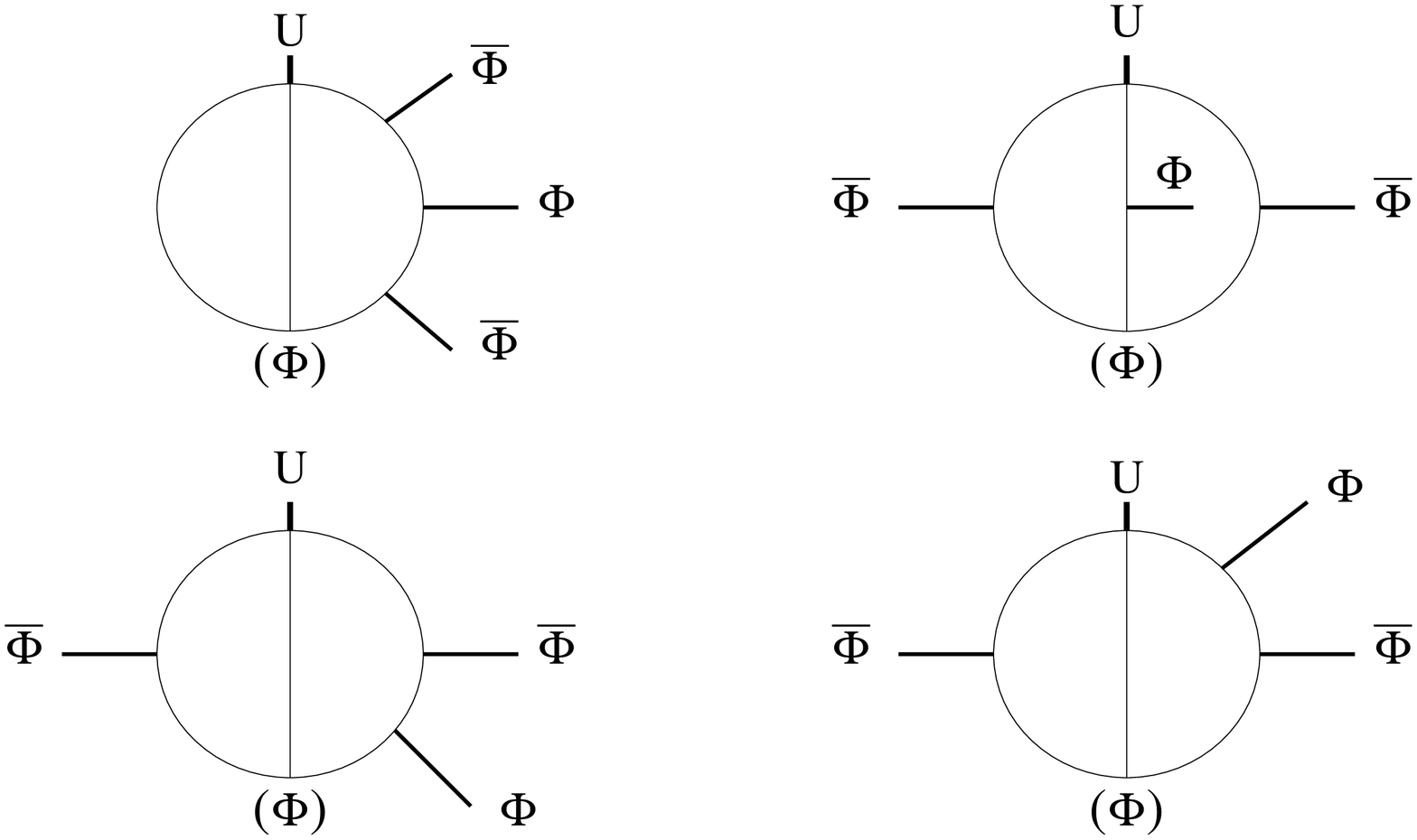}
\end{center}
\begin{center}
{\small{Figure 13: Two--loop contributions $U (D^2\Phi) \bar{\Phi}^2 \quad \rightarrow \quad C_{3}$ }}
\end{center}
\end{minipage}
\vskip 10pt
\begin{minipage}{\textwidth}
\begin{center}
\includegraphics[width=0.70\textwidth]{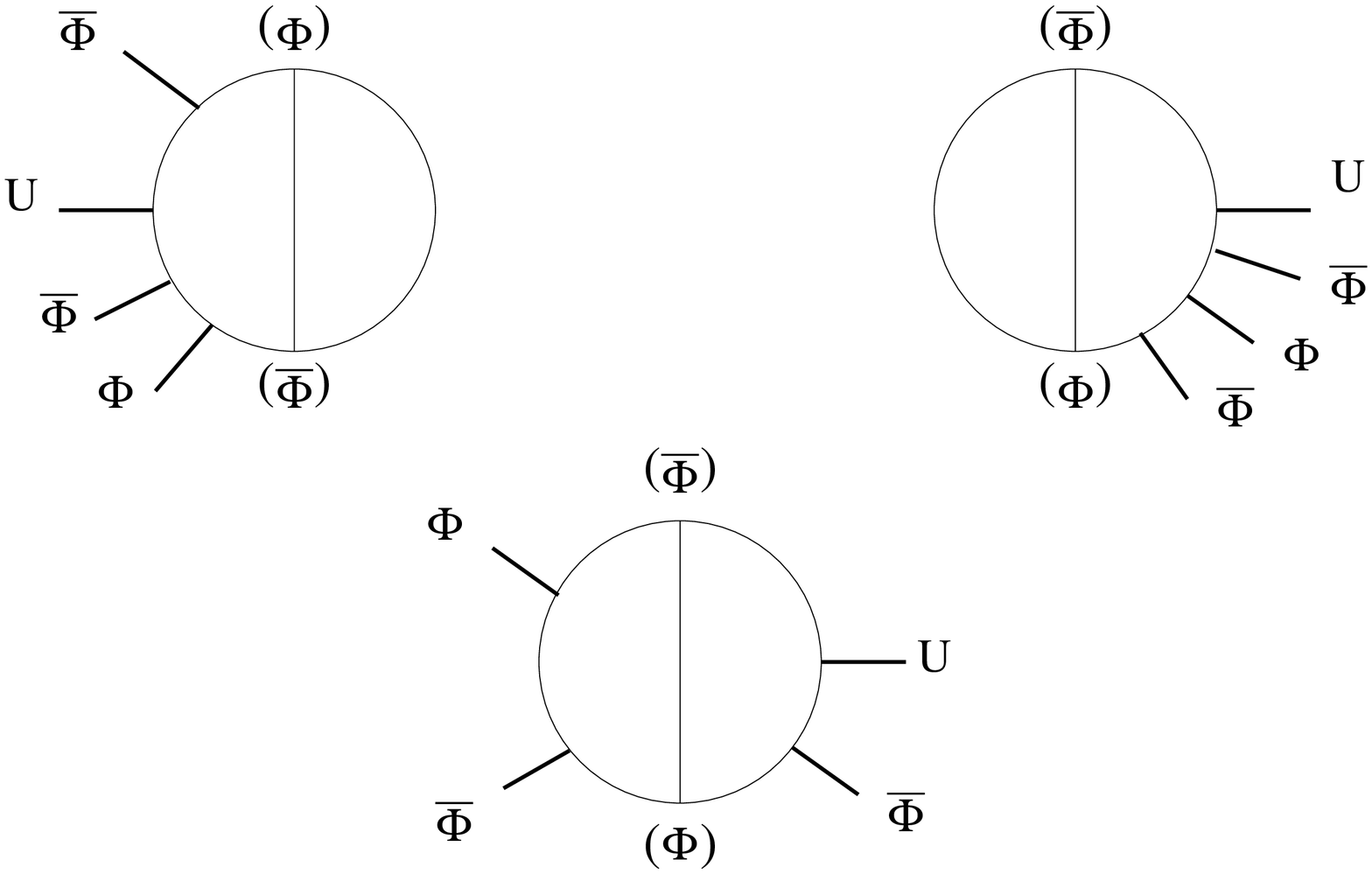}
\end{center}
\begin{center}
{\small{Figure 14: Two--loop contributions $U (D^2\Phi) \bar{\Phi}^2 \quad \rightarrow \quad \widetilde{C}_{3}$ }}
\end{center}
\end{minipage}
%---------- FIGURE END ------------

\vskip 20pt

Computing the combinatorial factors, we have the contributions
\bea
&&C_{1} ~:~  [4I_{2}]g^3 \bar{g}^2 \int d^8z U (D^2 \Phi)^3  ~\rightarrow~
- \frac{4}{\e^2} g^3 \bar{g}^2 C^2 \int d^4x ~F^3
\nonumber \\
&&C_{2} ~:~  [(8+16+8+8+8+16)I_{1}+16I_{2}]g^3 \bar{g}^2 \bar{m}
\int d^8z U (D^2 \Phi)^2 \bar{\Phi} \nonumber \\
&& \hspace{220pt} \rightarrow~ \left(- \frac{3}{\e^2}+\frac{2}{\e}\right)
16 g^3 \bar{g}^2 \bar{m} C^2 \int d^4x ~F^2~\bar{\phi}\nonumber \\
\nonumber \\
&&C_{3} ~:~  [(8+8+16+16)I_{1}]g^3 \bar{g}^2 \bar{m}^2 \int d^8z U (D^2\Phi)
\bar{\Phi}^2 \nonumber \\
&&\hspace{220pt} \rightarrow~  \left(- \frac{1}{\e^2}+\frac{1}{\e}\right)
24g^3 \bar{g}^2 \bar{m}^2 C^2 \int d^4x ~F~\bar{\phi}^2 \nonumber \\
&&\widetilde{C}_{3} ~:~  [(32+32+64)I_{1}]k g^2 \bar{g}^3 \bar{m}^2
\int d^8z U (D^2\Phi) \bar{\Phi}^2 \nonumber \\
&&\hspace{220pt} \rightarrow~
\left(- \frac{1}{\e^2}+\frac{1}{\e}\right) 64k_2 g^2 \bar{g}^3 \bar{m}^2 C^2
\int d^4x ~F~\bar{\phi}^2 \nonumber \\
&&~~~~~~~~
\eea

%%%%%%%%%%%%%%%%%%%%%%%%%%%%%%%%%%%%%%%%%%%%

\subsection{Four-point function}

For the 4-point functions the graphs drawn in Figs. 15,16 give divergent
contributions to the action

\vskip 18pt
\noindent
%---------- FIGURE TOP ------------
\begin{minipage}{\textwidth}
\begin{center}
\includegraphics[width=0.70\textwidth]{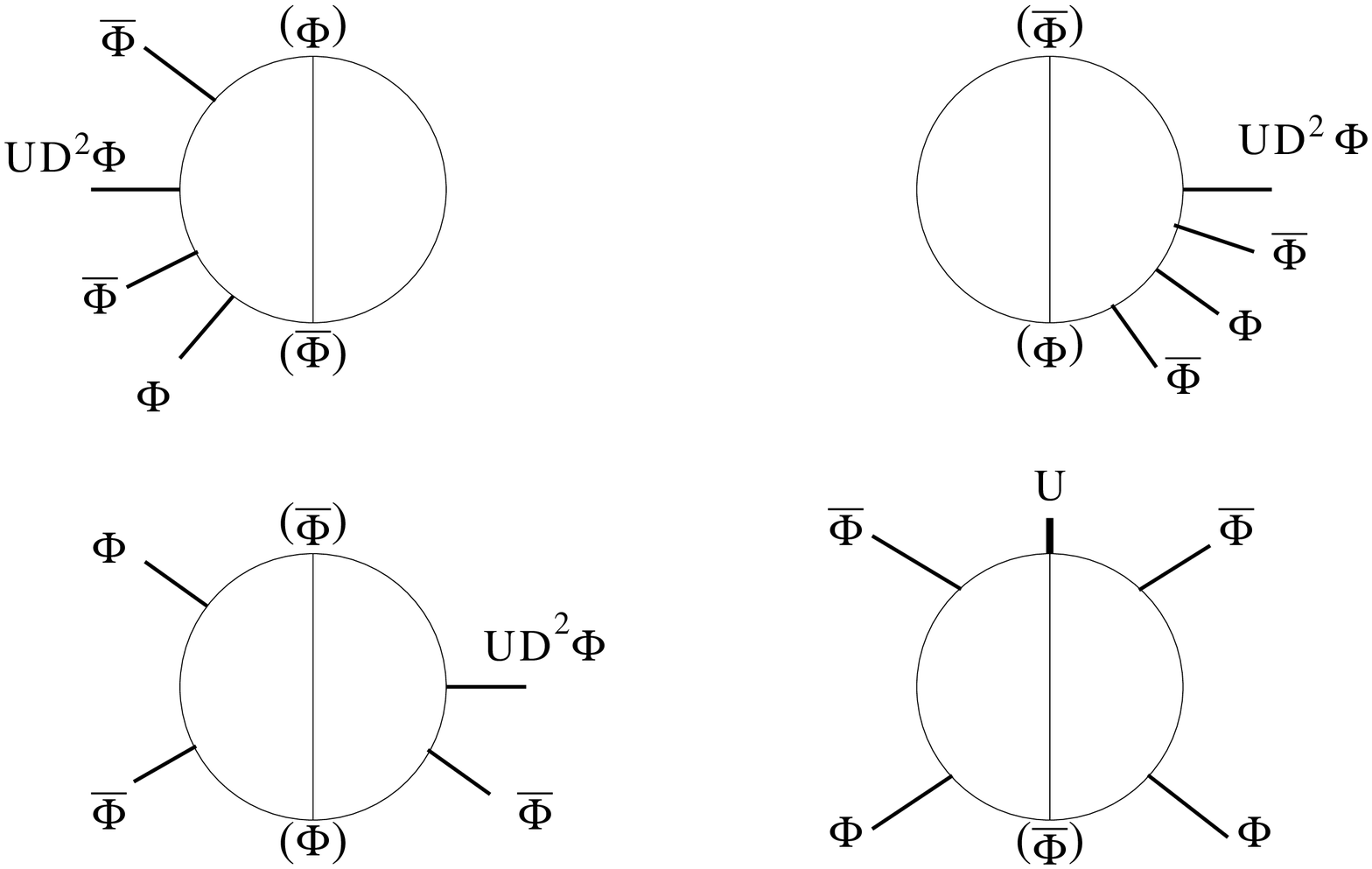}
\end{center}
\begin{center}
{\small{Figure 15: Two--loop contributions $U(D^2 \Phi)^2\bar{\Phi}^2 \quad \rightarrow \quad D_{1}$}}
\end{center}
\end{minipage}
\vskip 0.5cm
\begin{minipage}{\textwidth}
\begin{center}
\includegraphics[width=0.70\textwidth]{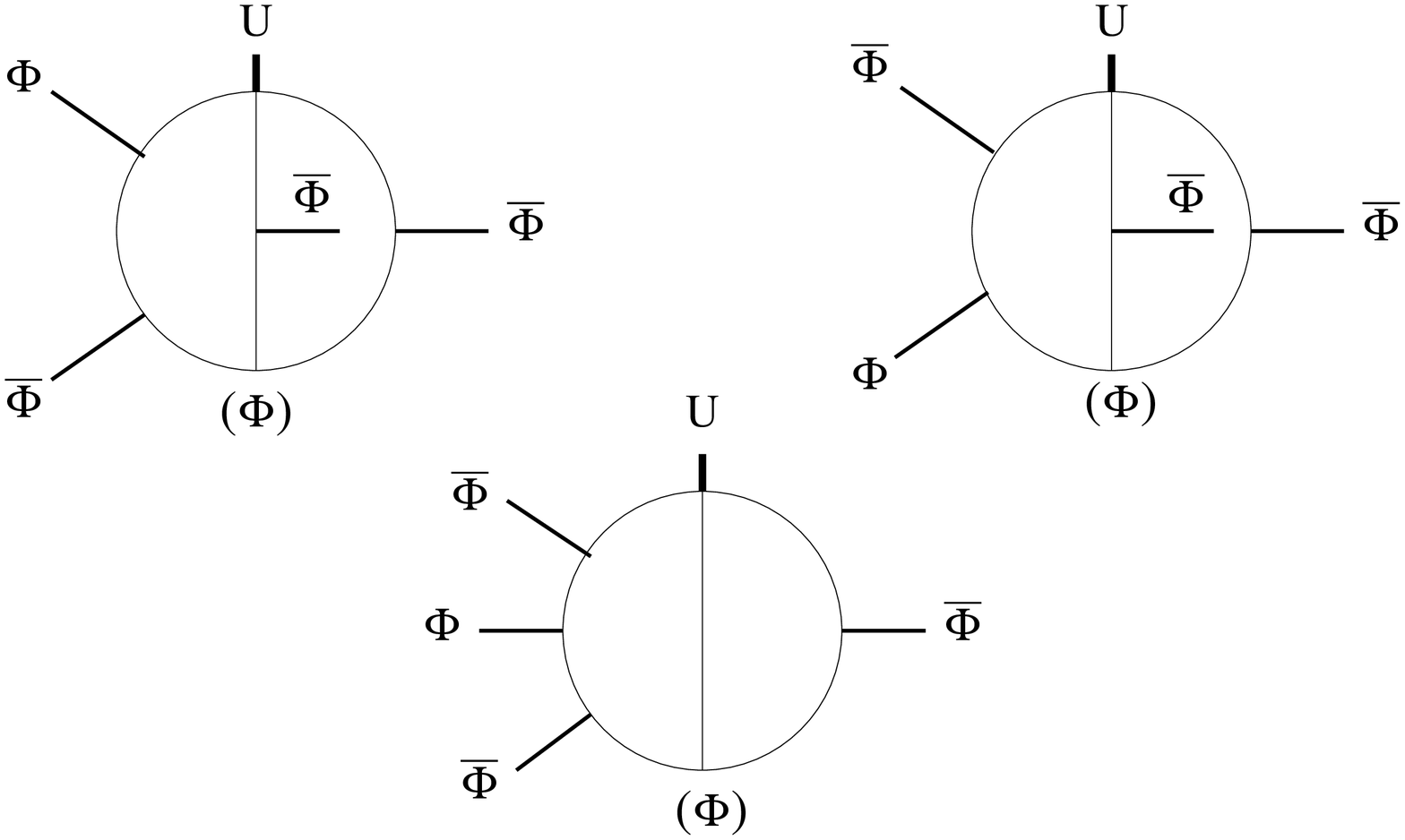}
\end{center}
\begin{center}
{\small{Figure 16: Two--loop contributions $U (D^2 \Phi) \bar{\Phi}^3 \quad \rightarrow \quad D_{2}$ }}
\end{center}
\end{minipage}
%---------- FIGURE END ------------

\vskip 20pt

Computing the combinatorial factors, the contributions are
\bea
&&D_{1} ~:~  [(16+16+32)I_{1}+16I_{2}]g^3 \bar{g}^3 \int d^8z U (D^2 \Phi)^2
\bar{\Phi}^2 \nonumber \\
&& \hspace{235pt} \rightarrow~\left(- \frac{3}{\e^2}+\frac{2}{\e}\right)
16g^3 \bar{g}^3  C^2 \int d^4x ~F^2~\bar{\phi}^2\nonumber \\
\nonumber \\
&&D_{2} ~:~  [(16+16+32)I_{1}]g^3 \bar{g}^3 \bar{m} \int d^8z U (D^2 \Phi)
\bar{\Phi}^3 ~\rightarrow~  \left(- \frac{1}{\e^2}+\frac{1}{\e} \right)
32 g^3 \bar{g}^3 \bar{m} C^2 \int d^4x ~F~\bar{\phi}^3\nonumber \\
\eea

%%%%%%%%%%%%%%%%%%%%%%%%%%%%%%%%%%%%%%%%%%%%

\subsection{Five-point function}

In this case we have only one divergent graph drawn in Fig. 17.
\vskip 18pt
\noindent
%---------- FIGURE TOP ------------
\begin{minipage}{\textwidth}
\begin{center}
\includegraphics[width=0.35\textwidth]{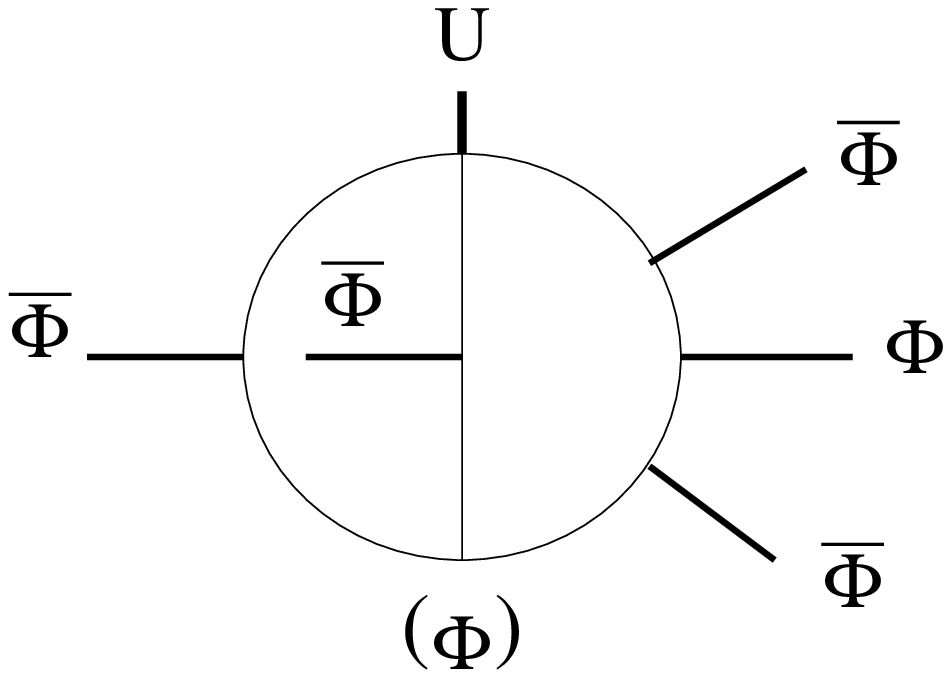}
\end{center}
\begin{center}
{\small{Figure 17: Two--loop contribution $ U (D^2 \Phi) \bar{\Phi}^4 \quad \rightarrow \quad E_{1}$}}
\end{center}
\end{minipage}
%---------- FIGURE END ------------

\vskip 20pt

Computing the combinatorial factors it gives
\begin{equation}
E_{1} ~:~  [32I_{1}]g^3 \bar{g}^4 \int d^8z U (D^2 \Phi) \bar{\Phi}^4
~\rightarrow~\left(- \frac{1}{\e^2}+\frac{1}{\e}\right) 16 g^3 \bar{g}^4 C^2
\int d^4x ~F~\bar{\phi}^4
\end{equation}
%%%%%%%%%%%%%%%%%%%%%%%%%%%%%%%%%%%%%%%%%%%%
\vskip 20pt

By power counting it is easy to discover that diagrams with more
than five external lines are always convergent.
Therefore, collecting all the results and inserting back the $4\pi$
factors, the total sum of two--loop divergences can be arranged in the
following expression
\beq
g^2 \bar{g} C^2 \int d^4x \Big[ a_1 F \bar{G} ~+~ a_2 F^2 \bar{G} ~+~
a_3 F \bar{G}^2 \Big] + g^2 C^2 \int d^4x  \Big[ a_4 F + a_5 F^2 + a_6 F^3
\Big]
\eeq
where the divergent coefficients are given by
\bea
&& a_1 = \frac{\bar{m}^2}{32\pi^4} (8 k_2 \bar{g} + g) \left(-\frac{1}{\e^2} +
\frac{1}{\e} \right)\qquad , \qquad a_2 = \frac{g \bar{g}}{16\pi^4}
\left( -\frac{3}{\e^2} + \frac{2}{\e} \right)
\nonumber \\
&& a_3 = \frac{g \bar{g}}{16\pi^4}\left(-\frac{1}{\e^2} + \frac{1}{\e}
\right)
\qquad ~~~~~~\qquad , \qquad~
a_4 =  \frac{\bar{m}^4}{256\pi^4} (g + 16 k_2 \bar{g})
\left(-\frac{1}{\e^2} + \frac{1}{\e} \right)
\nonumber \\
&& a_5 = \frac{\bar{g} \bar{m}^2}{64\pi^4}
\left[ \left(-\frac{3}{\e^2} + \frac{2}{\e} \right)g
- \frac{2}{\e^2} k_2 \bar{g} \right] \quad , \qquad
a_6 = - \frac{g \bar{g}^2}{64\pi^4} \frac{1}{\e^2}
\eea

Again, as explained in  Section 2, we can use the classical EOM so
as to replace the $\bar{G}$ factors in the first square bracket by factors of $F$.
There is one slight subtlety: as we have seen, replacing $F$
by ($\bar{m} \bar{\phi} + \bar{g} \bar{\phi}^2$) comes about essentially because one is
contracting the former factor with a
factor of $\bar{F}$, the contraction being equal to unity. However, if one is contracting
two such factors, $F^2$ with the corresponding factors in $\bar{F}^2$ a combinatorial
factor of 2 gets produced. Therefore, the correct replacement is $\bar{G}^2 \rightarrow
\frac{1}{2}F^2$.

Summing everything, our final result at two loops, in superspace language, reads
\bea
&&\int d^8z \Big\{
\frac{\bar{m}^4 g^2}{256 \pi^4} \left[ (g + 16 k_2 \bar{g})
\left( -\frac{1}{\epsilon^2} + \frac{1}{\epsilon}  \right)  \right] ~~U (D^2 \Phi)
\nonumber \\
&&\qquad ~+ \frac{\bar{m}^2 g^2}{64 \pi^4} \left[ g \bar{g}
\left( -\frac{5}{\epsilon^2} + \frac{4}{\epsilon}  \right) + k_2 \bar{g}^2
\left( - \frac{18}{\epsilon^2} + \frac{16}{\epsilon}  \right)  \right]
~~U (D^2 \Phi)^2  \nonumber \\
&&\qquad ~~+ \frac{g^3 \bar{g}^2}{64 \pi^4} \left[ -\frac{15}{\epsilon^2} +
\frac{10}{\epsilon} \right] ~~U (D^2 \Phi)^3\Big\}
\label{final}
\eea

\vskip 25pt
\noindent

\section{Renormalization and beta functions}

In this Section we perform the renormalization of the model at two loops and
compute the beta functions for the couplings. In particular, we
are interested in the renormalization group equation for
the nonanticommutation parameter $C_{\a \b}$.
In order to deal with
dimensionless quantities, we redefine $C^2 \rightarrow \g C^2$ with $\g$ the
dimensionless coupling subject to renormalization.

In dimensions $n = 4-2\e$ we define renormalized quantities as
\bea
&& \Phi = Z_{\Phi}^{-\frac12} \Phi_B \qquad \quad~~~~~~ , \qquad
\Phib = Z_{\Phib}^{-\frac12} \Phib_B
\nonumber \\
&& g = \mu^{-\e} Z_g^{-1} g_B \qquad \quad ~~~, \qquad
\bar{g} = \mu^{-\e} Z_{\bar{g}}^{-1} \bar{g}_B
\nonumber \\
&& k_1 = \mu^{\e} Z_{K_1}^{-1}(k_1)_B \qquad~~~ , \qquad k_2 = Z_{K_2}^{-1}
(k_2)_B
\nonumber \\
&& \g = Z_{\g}^{-1} \g_B \label{Z} \eea
where powers of the renormalization mass $\mu$ have been introduced
in order to deal with dimensionless renormalized couplings.

From the classical action written in terms of renormalized quantities plus
the divergent counterterms we can easily compute the $Z$ functions up to
two loops. 
We can immediately compute $Z_g$ and $Z_{\bar{g}}$
from $Z_{\Phi}$ by requiring that $g\Phi^3$ and $\bar{g} \bar{\Phi}^3$
 be not renormalized. If we set set $Z_{\Phi} = Z_{\Phib}$
(in this case there is no h.c. relation which forces this choice) we find
\bea
&& g~Z_g ~=~ g~\Big[ 1 + \left( \frac{3}{(4\pi)^2} g \bar{g} -
\frac{6}{(4\pi)^4}
g^2 \bar{g}^2 \right) \frac{1}{\epsilon} +\frac{27}{2(4\pi)^4} g^2 \bar{g}^2
\frac{1}{\epsilon^2} \Big] \equiv g + \frac{g_1}{\e} + \frac{g_2}{\e^2}
\nonumber \\
&& \bar{g}~Z_{\bar{g}} ~=~  \bar{g}~\Big[ 1 + \left( \frac{3}{(4\pi)^2}
g \bar{g} - \frac{6}{(4\pi)^4} g^2 \bar{g}^2 \right) \frac{1}{\epsilon} +
\frac{27}{2{(4\pi)^4}} g^2 \bar{g}^2
\frac{1}{\epsilon^2} \Big]\equiv \bar{g} + \frac{\bar{g}_1}{\e} +
\frac{\bar{g}_2}{\e^2}
\label{zeta1}
\eea
By writing the counterterms as in eq. (\ref{final}) we also find
\beq
\g~Z_{\g} ~=~ \g~ \Big[ 1 + \left( \frac{24}{(4\pi)^2} g \bar{g} -
\frac{240}{(4\pi)^4} g^2 \bar{g}^2 \right) \frac{1}{\epsilon} +
\frac{360}{(4\pi)^4} g^2 \bar{g}^2 \frac{1}{\epsilon^2} \Big]
\equiv \g + \frac{\g_1}{\e} + \frac{\g_2}{\e^2}
\label{zeta2}
\eeq
From the coefficients of the $1/\e$ pole we can compute the beta functions as
\bea
&& \b_g ~=~ - g \e - \left( 1 - g \frac{\pa}{\pa g} - \bar{g}
\frac{\pa}{\pa \bar{g}} \right) g_1
\nonumber \\
&& \b_{\bar{g}} ~=~ - \bar{g} \e - \left( 1 - g \frac{\pa}{\pa g} - \bar{g}
\frac{\pa}{\pa \bar{g}} \right) \bar{g}_1
\nonumber \\
&& \b_{\g} ~=~ \left( g \frac{\pa}{\pa g} + \bar{g}
\frac{\pa}{\pa \bar{g}} \right) \g_1
\eea
whereas from the poles equations
\bea
&&\left( 1 -  g \frac{\pa}{\pa g} - \bar{g} \frac{\pa}{\pa \bar{g}} \right) g_2
~=~ \frac{\pa g_1}{\pa g} \left( 1 -  g \frac{\pa}{\pa g} -
\bar{g} \frac{\pa}{\pa \bar{g}} \right) g_1
~+~ \frac{\pa g_1}{\pa \bar{g}} \left( 1 -  g \frac{\pa}{\pa g} -
\bar{g} \frac{\pa}{\pa \bar{g}} \right) \bar{g}_1
\nonumber \\
&&\left( 1 -  g \frac{\pa}{\pa g} - \bar{g} \frac{\pa}{\pa \bar{g}} \right)
\bar{g}_2
~=~ \frac{\pa \bar{g}_1}{\pa g} \left( 1 -  g \frac{\pa}{\pa g} -
\bar{g} \frac{\pa}{\pa \bar{g}} \right) g_1
~+~ \frac{\pa \bar{g}_1}{\pa \bar{g}} \left( 1 -  g \frac{\pa}{\pa g} -
\bar{g} \frac{\pa}{\pa \bar{g}} \right) \bar{g}_1
\nonumber \\
&&\left(g \frac{\pa}{\pa g} + \bar{g} \frac{\pa}{\pa \bar{g}} \right) \g_2
~=~ \frac{\pa \g_1}{\pa \g} \left( g \frac{\pa}{\pa g} +
\bar{g} \frac{\pa}{\pa \bar{g}} \right) \g_1
- \frac{\pa \g_1}{\pa g} \left( 1 - g \frac{\pa}{\pa g} -
\bar{g} \frac{\pa}{\pa \bar{g}} \right) g_1
\nonumber \\
&&~~~~~~~~~~~~~~~~~~~~~~~~~~~~~~~~~~~~~~-~
\frac{\pa \g_1}{\pa \bar{g}} \left( 1 -  g \frac{\pa}{\pa g} -
\bar{g} \frac{\pa}{\pa \bar{g}} \right) \bar{g}_1
\eea
we can make a nontrivial check of our calculations.

Inserting the explicit expressions (\ref{zeta1}, \ref{zeta2})
it is easy to prove that the pole equations are satisfied.
Concerning the beta functions, it is evident that the ones for the usual 
$g$, $\bar{g}$ couplings have the standard value
\bea
&&\b_{g} = - \epsilon g + \frac{3}{8 \pi^2} g^2 \bar{g} \left( 1 - \frac{g \bar{g}}{4 \pi^2} \right) \nonumber \\ 
&&\b_{\bar{g}} = - \epsilon \bar{g} + \frac{3}{8 \pi^2} g \bar{g}^2 \left( 1 - \frac{g \bar{g}}{4 \pi^2} \right)
\eea
whereas the beta function for $\g$ is given by
\beq
\b_{\g} ~=~ \frac{3}{\pi^2} \g g \bar{g} \left( 1-\frac{5}{4\pi^2}g \bar{g}
\right)
\label{beta}
\eeq
This result is independent of the particular choice $Z_{\Phi} =
Z_{\bar{\Phi}}$ we have made. However, since we are working in a theory with several
coupling constants scheme-dependence may appear already at the two-loop level and
undermine the reliability of the second term. Nonetheless, it is interesting to note
that a nontrivial fixed point for the $C^2$ coupling may exist.

\section{Conclusions}

In this paper we have studied perturbatively the WZ model defined on
a nonanticommutative (NAC) superspace where the $\theta$ variables are not
ordinary Grassmann variables but satisfy a Clifford algebra. Consequently, the
usual superspace WZ action is augmented by a term  which cannot be written
 immediately as a superspace integral of superfields.
In order to apply standard superspace techniques in the course of our calculations,
we have chosen to describe the effects of the NAC (the additional $F^3$ term in
eq. (\ref{components}))
through the introduction of a
{\it spurion} superfield. This is a constant superfield proportional to $C^2$,
 the square of the NAC parameter, with only a nonvanishing  highest
component.
We have performed a systematic analysis of all the divergent contributions
up to two loops and obtained the following results:

\begin{itemize}
\item
Up to this order divergent diagrams always contain only one insertion of
the spurion. This means that divergent contributions to the effective
action proportional to higher powers of $C^2$ are absent.
\item
At one loop a divergence appears proportional to $F^2$ in agreement with
results
found in \cite{Rey, TY}. In order to deal with a renormalizable model we
must start with a classical action containing additional $F$, $F^2$ terms
(we choose to add also the linear term because from $F^2$ one naturally starts
producing tadpoles).
Up to second order in perturbation theory we
computed all the divergent diagrams with vertices $F^2$ and $F^3$ and showed
that no new divergent structures  emerge. Up to this order
the theory is renormalizable.
\item
Up to two loops the new divergences which arise due to the
$F^3$ vertex are still logarithmic as in the ordinary, supersymmetric case,
thus proving that
nonanticommutativity induces a soft--breaking of supersymmetry.
\item
We have studied the renormalization of the theory and computed the beta
functions. Even if at two loops we expect these functions to be affected
by scheme dependence it is anyway interesting to see the structure of
the beta function associated to the NAC parameter.
As appears from our result (\ref{beta}) nontrivial fixed points
might exist.

\end{itemize}

The appearance of $F^2$ divergent terms might lead to the conclusion that the star product gets deformed at the quantum level. On general grounds this shouldn't happen since an alternative way to perform calculations would be to keep the star product implicit and implement in superspace the technologies developed for dealing with NC bosonic coordinates. Indeed, the authors of ref. \cite{Rey2} give a general argument to prove that suitable resummations of this and others terms in the effective potential can be rewritten in terms of star product.

Our results confirm the generalized non--renormalization theorem formulated
in \cite{Rey}. In our language the general structure of the effective
action reads 
\beq
\G[\Phi, \Phib] ~=~ \sum_n \int \prod_{j=1}^n d^4x_j
 \int d^2\theta d^2 \thb G_n(x_1, \cdots , x_n, U, D^2\bar{D}^2 U)
F_1(x_1, \theta,\thb)
\cdots F_n(x_1, \theta,\thb)
\eeq
where $G_n$ may depend on $U$ linearly and on an arbitrary polynomial in $(D^2\bar{D}^2U)$.
Up to two loops the divergent part of the functions $G_n$ depend only on $U$ and not on its derivatives. It would be interesting to find an argument to prove
that this statement remains true at any order in perturbation theory.
It would  also be interesting to study the model at higher loops and find
a general argument for renormalizability at every order.

Finally, we note that our approach can be easily extended to the case
of matter coupled to gauge fields.

\vskip 30pt
{\bf Acknowledgments} This work has been supported in
part by INFN, MURST and the European Commission RTN program
HPRN--CT--2000--00131, in which S.P. and A.R. are associated to
the University of Padova. The work of M.T.G. is supported by NSF Grant No. PHY-0070475 and by NSERC Grant No. 204540.

\newpage
\appendix
\section{Divergent diagrams from $U$-insertions }

In this Appendix we use power counting to prove that at one and two
loop order
only diagrams with at most one insertion of $U$ vertices can be potentially
divergent.

We consider the most general one--loop graph as in Fig. A1
\vskip 18pt
\noindent
%---------- FIGURE TOP ------------
\begin{minipage}{\textwidth}
\begin{center}
\includegraphics[width=0.70\textwidth]{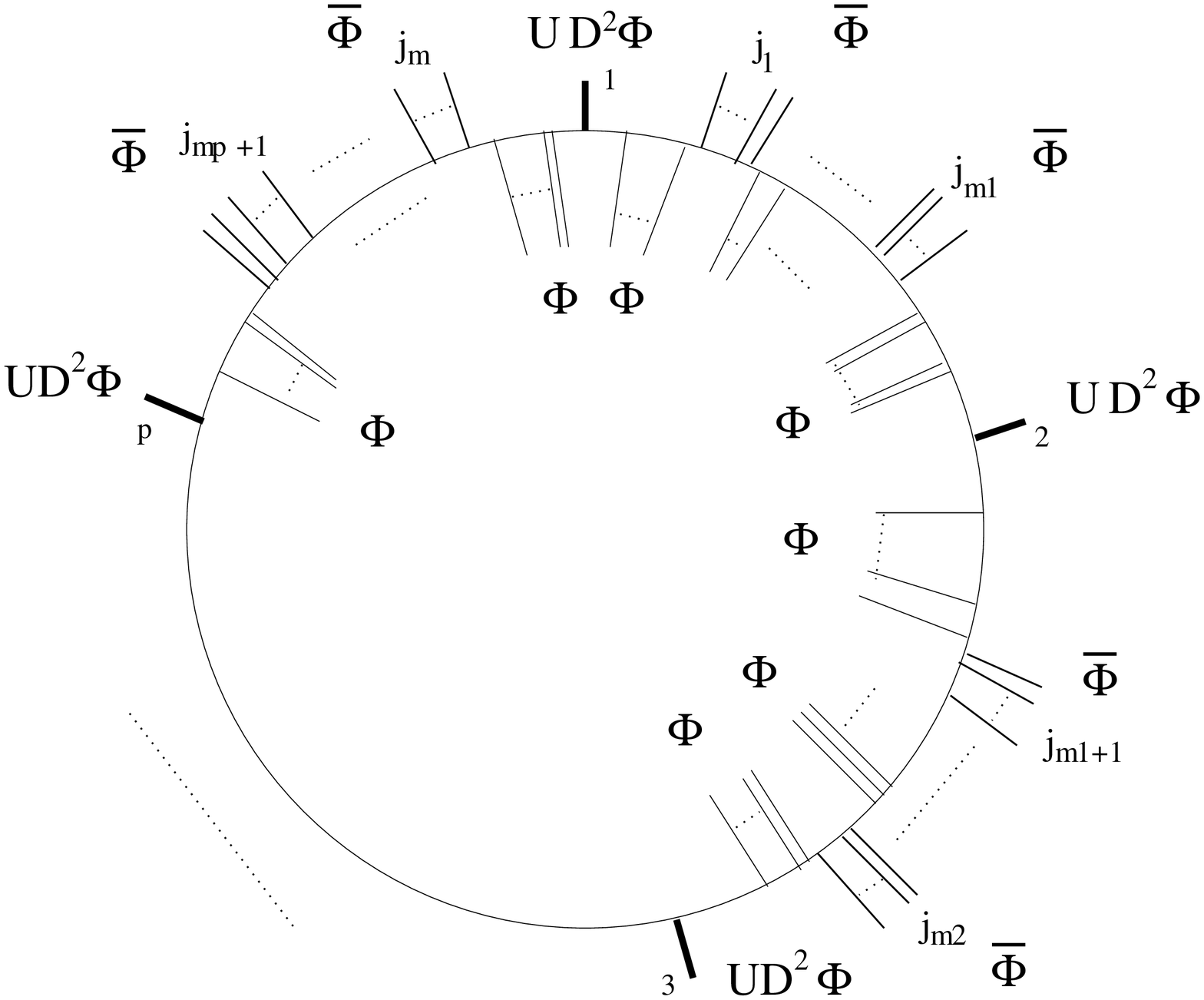}
\end{center}
\begin{center}
{\small{Figure A1
}}
\end{center}
\end{minipage}
%---------- FIGURE END ------------
\vskip 20pt

This diagram contains  $p$ $U(D^2 \Phi)^3$ vertices, $k$ $\bar{\Phi}^3$
vertices organized in $m$ groups and $n-k$ vertices $\Phi^3$ distributed
among the $\Phib$ groups.
The total number of external legs is $n+p$ and $k \geq m$.
Moreover, we indicate with
$j_{i}$ the number of adjacent $\bar{\Phi}^3$ vertices in the
$i^{th}$ group. According to the number of $\Phi$ vertices between two $\Phib$
blocks the $U$ superfield is inserted between two $\Phib$'s, two $\Phi$'s
or one $\Phi$ and one $\Phib$.

To study possible divergent configurations we evaluate the mass dimensions
of the corresponding integrals once the $D$--algebra has been performed.
The first step is to compute the number of $D^2$, $\bar{D}^2$ derivatives
and the number of propagators initially present in the graph. Then we
perform $D$-algebra and look for the most divergent
configurations which are produced. They may occur when
we generate momentum factors through the algebraic relations
\bea \label{D--momentum}
&&D^2\bar{D}^2 D^2 = \Box D^2 \qquad \qquad \bar{D}^2 D^2 \bar{D}^2 =
\Box \bar{D}^2 \nonumber \\
&&~[ D^{\alpha}, \bar{D}^2 ] = i \partial^{\alpha \dot{\alpha}}
\bar{D}_{\dot{\alpha}} ~~~ \qquad [ \bar{D}^{\dot{\alpha}}, D^2 ] =
i \partial^{\dot{\alpha} \alpha} D_{\alpha}
\eea

From the Feynman rules described in the text it is easy to determine
that the initial number of $D^2$ and $\bar{D}^2$ is:
\begin{itemize}
\item
$D^2$ derivatives:
\begin{itemize}\item[-]2 for every $U$-vertex
                     \item[-]1 for every $\bar{\Phi}$-vertex
                     \item[-]1 for every propagator $\langle \Phi \Phi
                    \rangle$
\end{itemize}
\item
$\bar{D^2}$ derivatives:
\begin{itemize}\item[-]2 for every $U$-vertex
                           \item[-]1 for every $\Phi$-vertex
                           \item[-]1 for every propagator $\langle
                           \bar{\Phi}\bar{\Phi} \rangle$
\end{itemize}

\item
Propagators:
\begin{itemize}
\item
$\langle \bar{\Phi}\bar{\Phi} \rangle$ propagators: $\sum_{i=1}^m (j_i - 1) =
k-m$
\item
$\langle \Phi \Phib \rangle$ propagators: $2m$
\item
$\langle \Phi \Phi \rangle$ propagators: $(n+p)-(k-m)-2m$
\end{itemize}
\end{itemize}

The total number of derivatives is then
\bea
&& D^2 : n-m+3p                 \nonumber \\
&& \bar{D^2} : n-m+2p                   \nonumber \eea The $D$-algebra gives a
nonzero result when only  one pair $D^2\bar{D}^2$ survives inside the loop. We
can get rid of  additional covariant derivatives by integration by parts at the
vertices, either moving derivatives onto external legs or onto internal lines
where we can use then the identities (\ref{D--momentum}). The most divergent
configuration is the one where the maximum number of covariant derivatives
remain inside the diagram and combine into momentum factors. We study this case
in detail.

We recall that the $U$ superfield has only the $\theta^2\bar{\theta}^2$
component different from zero. This forces us to move one pair $D^2 \bar{D}^2$
onto each of $(p-1)$ external $U$'s in order to obtain a final nonvanishing
expression, so decreasing the number of $D^2$ and $\bar{D}^2$ inside the loop
by a factor $(p-1)$ (on the remaining $U$ superfield the $d^2 \theta d^2
\bar{\theta}$ integration will act). Taking into account that if the number of
$D$'s inside the loop is different from the number of $\bar{D}$'s the
$D$-algebra gives a vanishing contribution, we have to integrate by parts at
least $p$ $D^2$ derivatives onto external $\Phi$ superfields. This implies that
at least $p$ external $\Phi$-legs have to be present, i.e. \beq n-k \geqslant p
\eeq Now we can move the remaining derivatives inside the diagram and produce
momentum factors as in (\ref{D--momentum}).

In conclusion, considering that we have to end up with one pair $D^2 \bar{D^2}$
inside the loop, the $D$-algebra can produce at most a factor
\begin{equation} \label{numerat}
\Box^{n-m+p}
\end{equation}
This contribution enters the momentum integrals together with the contributions
from the propagators
\begin{itemize}
\item[]$\Box^{-2m}$~~~~~~~~~ from the $2m$  $\langle \Phi \bar{\Phi} \rangle$
propagators
\item[]$\Box^{-2(n+p-2m)}$ ~from the $\langle \Phi \Phi \rangle$ and
$\langle \bar{\Phi}\bar{\Phi} \rangle$ propagators
\end{itemize}
Therefore, the corresponding momentum integral ($\Box \rightarrow -q^2$)
behaves, in the UV region, as
\beq
\int d^4 q~\frac{1}{(q^2)^{[n-m+p]}}
\eeq
By power counting this integral is divergent if
 \begin{equation}
n \leqslant 2 + m - p
\end{equation}
This is consistent with the previous constraints
$n \geqslant k + p \geqslant m + p$ if and only if $p \leqslant 1$.
This concludes our proof that one--loop diagrams with more than one
$U$-insertion cannot be divergent.

We note that the same conclusion can be easily reached if
insertions of the $U(D^2\Phi)^2$ are considered, since the $D$-algebra is
identical.

At this point one can use the same kind of analysis to select the diagrams in Figs.
2,3 as the only divergent diagrams from one--loop graphs with one $U$-insertion.

%%%%%%%%%%%%%%%%%%%%%%%%%%%%%%%%%%%%%

\section{Two--loop divergent diagrams from $U$-insertions}

We now move to two loops  and show that again the superficially divergent
diagrams can have at most one insertion of the $U(D^2\Phi)^3$ vertex. The
procedure is the same as in the one--loop case: we look for the most divergent
configurations and we show that, since we have to move factors of $\bar{D}^2$
onto $U$-fields to get a nonvanishing expression, if there is more than one
$U$-insertion, the momentum factor that we are left with from $D$-algebra is
not sufficient to give a divergent term. Unlike the one-loop case, at this
order there are topologically different configurations. Schematically, these
are represented in Figs. B1--B6, where it is understood that there are
$p\geqslant 1$ insertions of $U$-terms among $k$ $\bar{\Phi}^3$ vertices
organized in $m$ groups ($k \geq m$) separated by an arbitrary number of
$\Phi^3$ vertices whose total number is $(n-k)$.

\vskip 18pt
\noindent
%---------- FIGURE TOP ------------
\begin{minipage}{\textwidth}
\begin{center}
\includegraphics[width=0.40\textwidth]{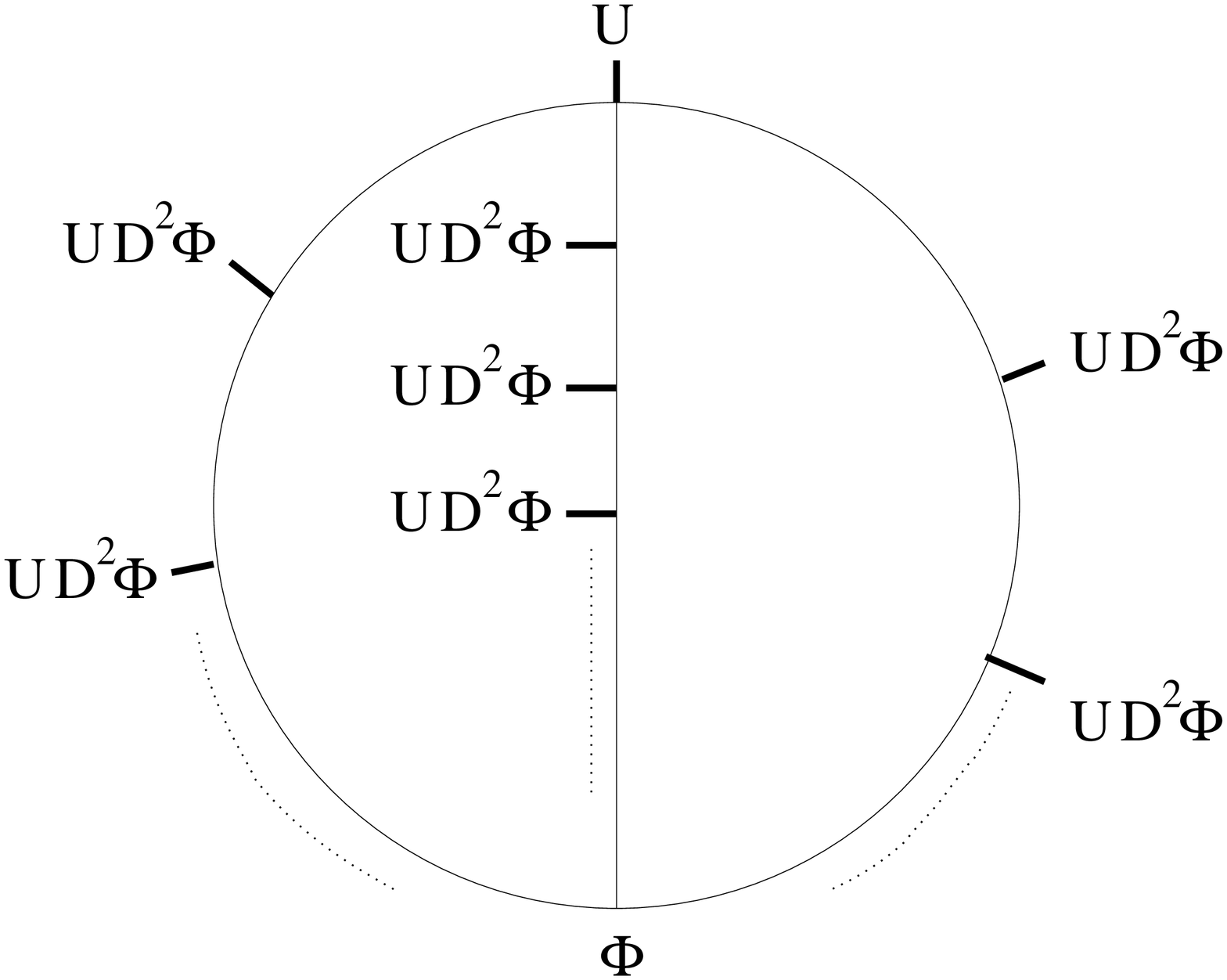}
\end{center}
\begin{center}
{\small{Figure B1}}
\end{center}
\end{minipage}
%---------- FIGURE END ------------
\vskip 20pt

We analyze the different topologies separately.

B1) In this case, when counting the number of $D^2$, $\bar{D}^2$ derivatives,
we have to take into account that the internal $U$-vertex brings three
$D^2$ and three $\bar{D}^2$, the internal $\Phi^3$ vertex gives two
$\bar{D}^2$. Counting $D^2$'s and $\bar{D}^2$'s from propagators
\bea
\langle \Phi \Phi \rangle ~~\textrm{propagators}
&&\rightarrow~
[(n+p+2)- (k-m) -2m]~D^2  \nonumber \\
\langle \bar{\Phi} \bar{\Phi} \rangle ~~\textrm{propagators}
&&\rightarrow~  (k-m)~\bar{D}^2  \nonumber
\eea
and from the vertices with external legs, we are led to
\bea
&&D^2 : n-m+3+3p \nonumber \\
&&\bar{D}^2 : n-m+3+2p \eea We can proceed exactly as in the one--loop case
with the only difference that now the $D$-algebra ends when two pairs
$D^2\bar{D}^2$ remain inside the graph (one for each loop). As before, we need
at least $p$ external $\Phi$ superfields to get rid of the extra $p$ $D^2$
derivatives and this imposes the extra constraint
\begin{equation}
n-k \geqslant p
\end{equation}
Finally, from $D$-algebra at most a momentum factor
\begin{equation}
\Box^{n-m+2+p}
\end{equation}
can be produced. Taking into account the momentum factors from the propagators
the most potentially divergent diagram has dimension
$8 -2(n-m+2+p)$ and it diverges if
 \begin{equation}
n \leqslant 2 + m - p
\end{equation}
Toghether with the other constraint $n \geqslant m+p$
this condition necessarily implies $p \leqslant 1$.

%%%%%%%%%%%%%%%%%%%%%%%%%%%%%%%%%%%%%%%%%%%%%%%%%%%%%%%
%%%%%%%%%%%%%%%%%%%%%%%%%%%%%%%%%%%%%%%%%%%%%%%%%%%%%%%
\vskip 18pt
\noindent
%---------- FIGURE TOP ------------
\begin{minipage}{\textwidth}
\begin{center}
\includegraphics[width=0.40\textwidth]{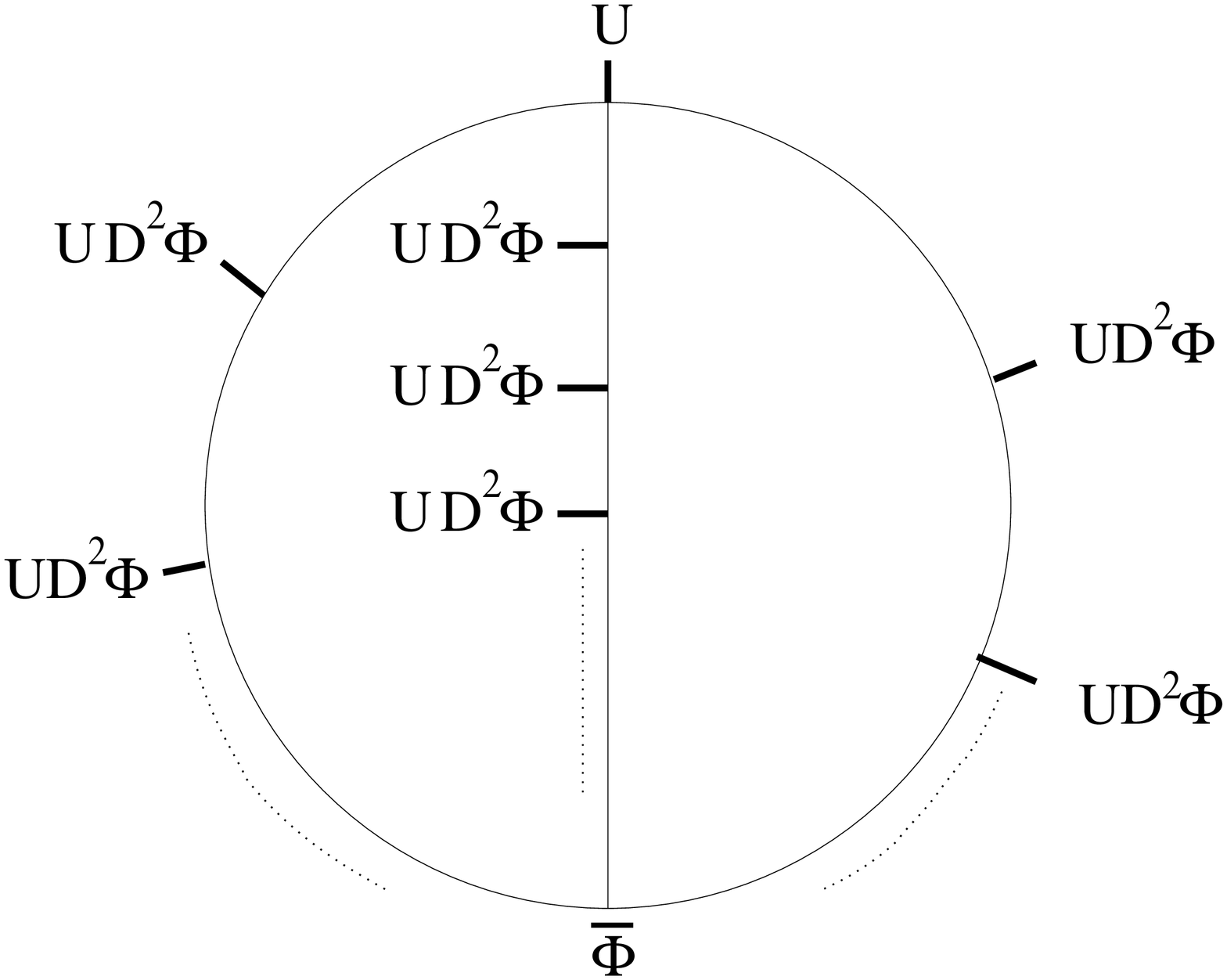}
\end{center}
\begin{center}
{\small{Figure B2}}
\end{center}
\end{minipage}
%---------- FIGURE END ------------
\vskip 20pt

B2) In this case the $U$-vertex with three internal legs gives
three $D^2$ and three $\bar{D}^2$ and the internal $\bar{\Phi}^3$ vertex
gives two $D^2$. The counting of derivatives from vertices with external
legs is the same as in the one--loop case. Moreover, we have a number of
derivatives from the chiral and antichiral propagators whose number is
\bea
\langle \Phi \Phi \rangle ~~\textrm{propagators}&&
\rightarrow~ [(n+p+2)- (k-m) -(2m-f+3))] D^2 \nonumber \\
\langle \bar{\Phi} \bar{\Phi} \rangle ~~\textrm{propagators}
&& \rightarrow~ (k-m+f) \bar{D}^2 \nonumber
\eea
where $f=0,1,2,3$ counts the number of $\bar{\Phi}^3$ vertices directly
connected to the internal $\bar{\Phi}$ vertex.
The total number of covariant derivatives is then
\bea
&&D^2 : n-m+3p+2+f \nonumber \\
&&\bar{D}^2 : n-m+2p+1+f
\eea
In the most divergent configuration, i.e. the one where the maximal
number of covariant derivatives remain inside the loop,
the $D$-algebra produces a momentum factor \begin{equation}
\Box^{n-m+p+f}
\end{equation}
subject to the constraint
\begin{equation}
n \geqslant k+p+1 \geqslant m+p+1
\end{equation}
as follows from the requirement to have at least $(p+1)$ external $\Phi$'s
to integrate out the additional $(p+1)$ $D^2$ derivatives.

Since the propagators give a power
\begin{equation}
\Box^{-[2n-2m+2p+2f+1]}
\end{equation}
the dimension of the corresponding integral is $8-(n-m+f+1+p)$.
It is then UV divergent if
\begin{equation}
n \leqslant 3 + m -f - p
\leqslant 3 + m - p
\end{equation}
Together with $n \geqslant m + p + 1$ it gives $p \leqslant 1$.

%%%%%%%%%%%%%%%%%%%%%%%%%%%%%%%%%%%%%%%%%%%%%%%%%%%%%%%%
%%%%%%%%%%%%%%%%%%%%%%%%%%%%%%%%%%%%%%%%%%%%%%%%%%%%%%%%
\vskip 18pt
\noindent
%---------- FIGURE TOP ------------
\begin{minipage}{\textwidth}
\begin{center}
\includegraphics[width=0.40\textwidth]{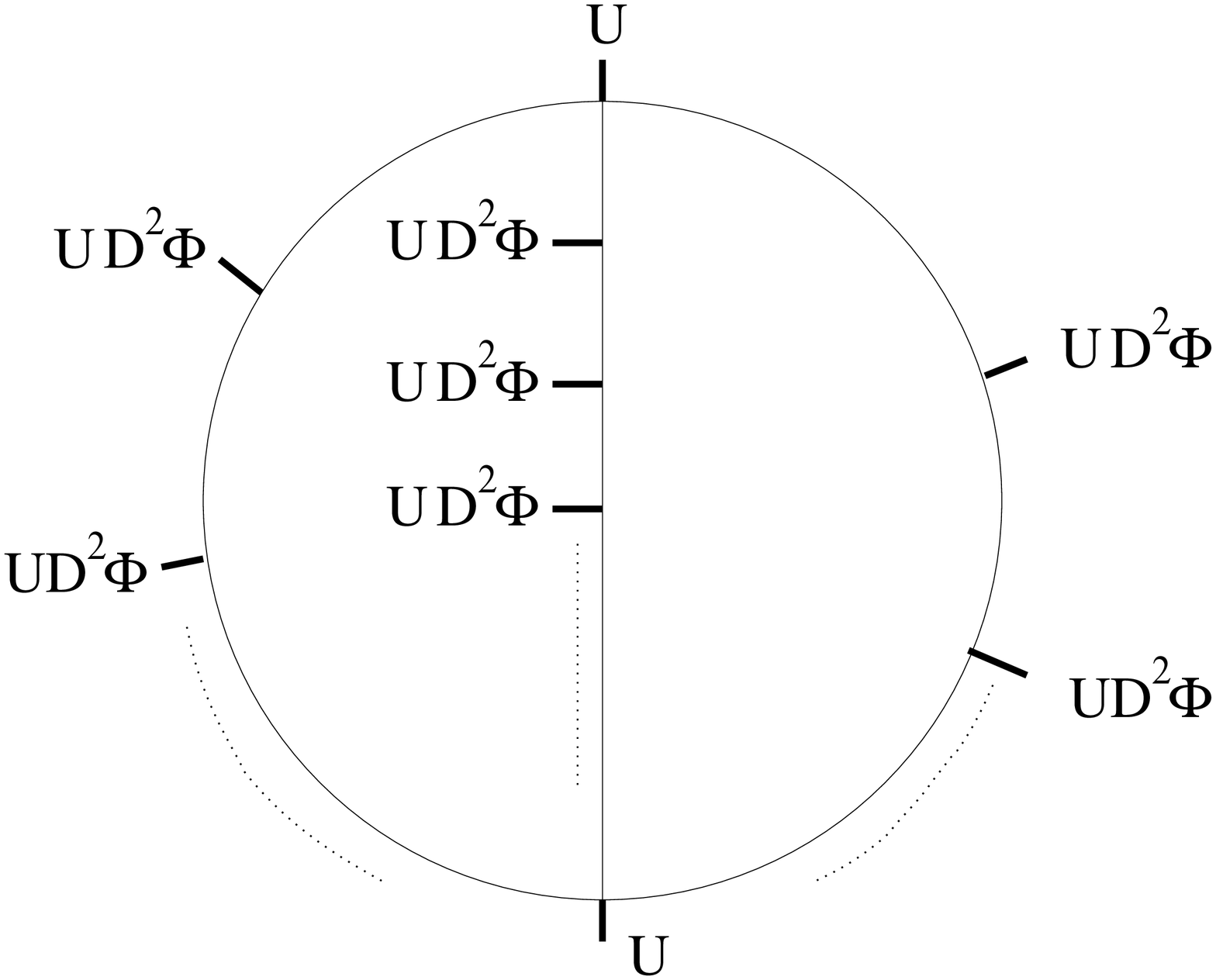}
\end{center}
\begin{center}
{\small{Figure B3}}
\end{center}
\end{minipage}
%---------- FIGURE END ------------
\vskip 20pt

B3) This configuration exists only when $p \geqslant 2$.
Since the two $U$-vertices with three internal legs gives six $D^2$ and six
$\bar{D}^2$ and
\bea
\langle \Phi \Phi \rangle ~~\textrm{propagators}&&
\rightarrow~ [(n+p+1)- (k-m) -2m] D^2 \nonumber \\
\langle \bar{\Phi} \bar{\Phi} \rangle ~~\textrm{propagators}&&
\rightarrow~ (k-m) \bar{D}^2\nonumber
\eea
the total number of covariant derivatives is
\bea
&&D^2 : n-m+3p+3 \nonumber \\
&&\bar{D}^2 : n-m+2p+2
\eea
In the most divergent configuration, the $D$-algebra produces a momentum
factor
\begin{equation}
\Box^{n-m+p+1}
\end{equation}
and the condition
\begin{equation}
n \geqslant k+p+1 \geqslant m+p+1
\end{equation}
Taking into account the factors from the propagators
\begin{equation}
\Box^{-[2n-2m+2p+2]}
\end{equation}
the corresponding momentum integral has dimension $8-(n-m+1+p)$ and
diverges if
 \begin{equation}
n \leqslant 3 + m - p
\end{equation}
Since $n \geqslant m + p + 1$, we find $p \leqslant 1$ which is incompatible
with the initial assumption $p \geqslant 2$. Therefore, there are no divergent
diagrams with this topology.

%%%%%%%%%%%%%%%%%%%%%%%%%%%%%%%%%%%%%%%%%%%%%%%%%%%%%%%%%
%%%%%%%%%%%%%%%%%%%%%%%%%%%%%%%%%%%%%%%%%%%%%%%%%%%%%%%%%
\vskip 18pt
\noindent
%---------- FIGURE TOP ------------
\begin{minipage}{\textwidth}
\begin{center}
\includegraphics[width=0.40\textwidth]{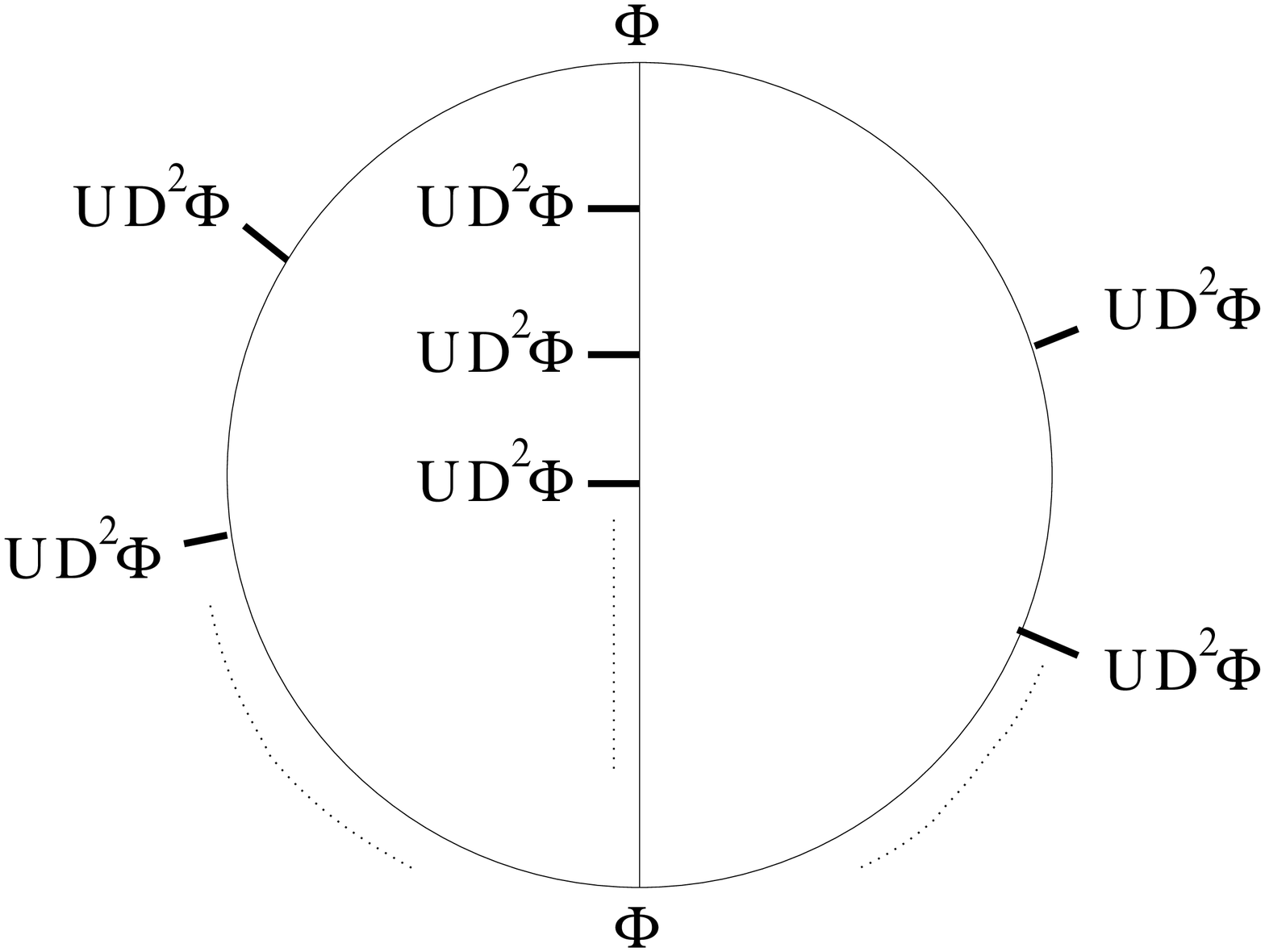}
\end{center}
\begin{center}
{\small{Figure B4}}
\end{center}
\end{minipage}
%---------- FIGURE END ------------
\vskip 20pt

B4) In this case, since two internal $\Phi^3$ vertices give four
$\bar{D}^2$ and
\bea
\langle \Phi \Phi \rangle ~~\textrm{propagators}&&
\rightarrow
[(n+p+3)- (k-m) -2m] D^2 \nonumber \\
\langle \bar{\Phi} \bar{\Phi} \rangle ~~\textrm{propagators}&&
\rightarrow (k-m) \bar{D}^2, \nonumber
\eea
the total number of covariant derivatives is
\bea
&&D^2 : n-m+3+3p \nonumber \\
&&\bar{D}^2 : n-m+4+2p \eea We first analyze the $p=1$ case which is the only
case with an equal number of $D^2$ and $\bar{D}^2$ from the beginning. The most
convenient way to perform $D$--algebra is to pull out a $D^2$ onto the
$U$-vertex by integration by parts. As a consequence, we have to pull out one
$\bar{D}^2$ in order to restore the equal number of chiral and antichiral
derivatives. In the most divergent configuration, completion of $D$-algebra
produces a momentum factor
\begin{equation}
\Box^{n-m+3}
\end{equation}
The propagators give a factor
\begin{equation}
\Box^{-[2n-2m+8]}
\end{equation}
and the corresponding momentum integral would be divergent if
$n \leqslant m -1 $ wich is obviously impossible.

In the general case with $p \geqslant 2$, the most divergent
configuration is realized when $D$-algebra produces a momentum factor
\begin{equation}
\Box^{n-m+3+p}
\end{equation}
and the condition
\begin{equation}
n \geqslant m + p -1
\end{equation}
is satisfied.
Since the propagators give a factor
\begin{equation}
\Box^{-[2n-2m+6+2p]}
\end{equation}
the corresponding momentum integral has dimension $8 - 2(n-m+3+p)$
and it is divergent if
\begin{equation}
n \leqslant 1 + m - p
\end{equation}
Together with $n \geqslant m + p -1$ it implies $p \leqslant 1$, in contrast
with our initial assumption $p \geqslant 2$.

In conclusion we find that the present topology of graphs can never
produce UV divergences.

%%%%%%%%%%%%%%%%%%%%%%%%%%%%%%%%%%%%%%%%%%%%%%%%%%%%%%%%%%%
%%%%%%%%%%%%%%%%%%%%%%%%%%%%%%%%%%%%%%%%%%%%%%%%%%%%%%%%%%%
\vskip 18pt
\noindent
%---------- FIGURE TOP ------------
\begin{minipage}{\textwidth}
\begin{center}
\includegraphics[width=0.40\textwidth]{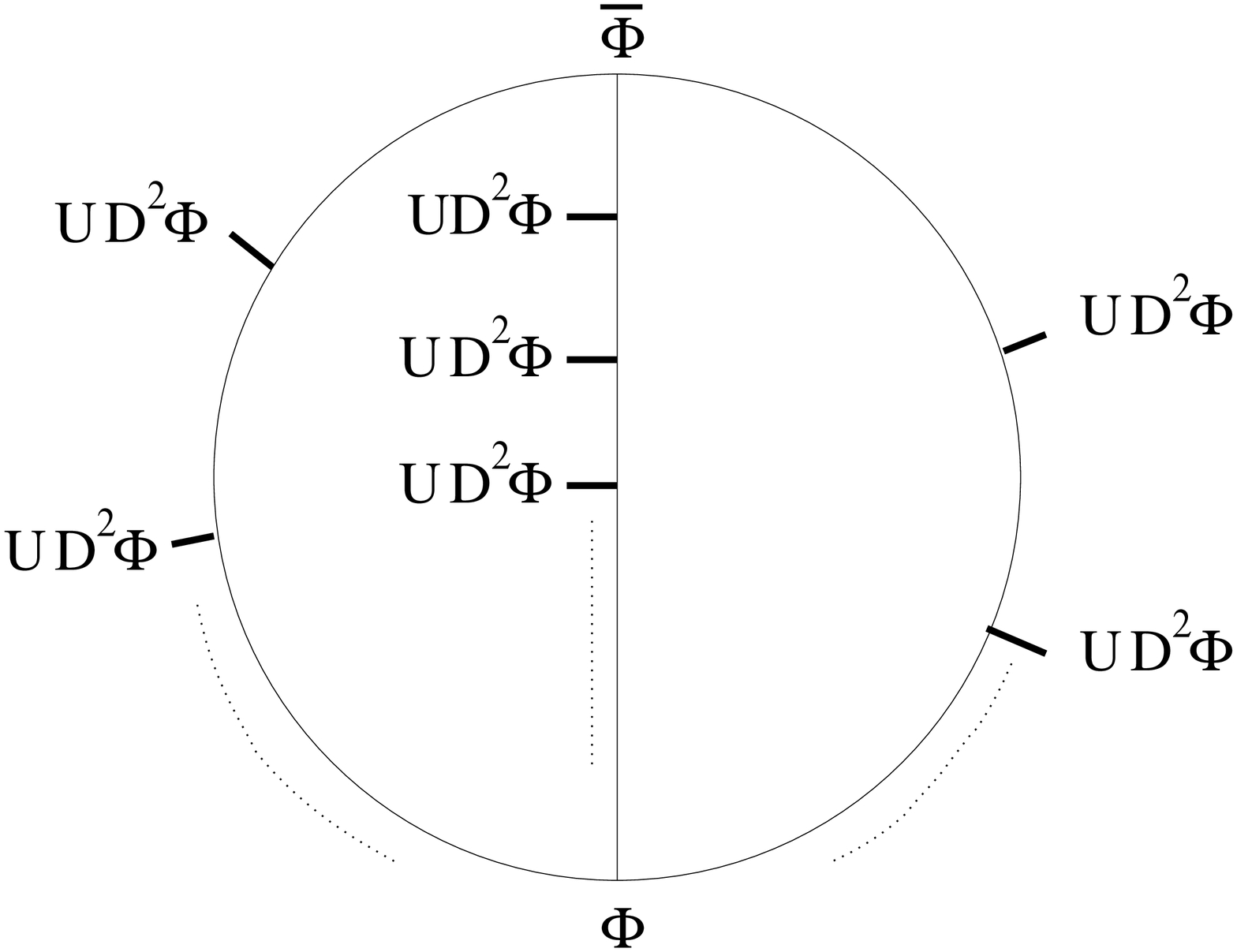}
\end{center}
\begin{center}
{\small{Figure B5}}
\end{center}
\end{minipage}
%---------- FIGURE END ------------
\vskip 20pt

B5) This case can be treated like the B2 case by introducing the $f$ parameter
which counts the number of $\Phib$ vertices directly connected to the
internal $\Phib^3$. In this case
since an internal $\bar{\Phi}^3$ vertex gives two $D^2$,
an internal $\Phi^3$ vertex gives two $\bar{D}^2$,
\bea
\langle \Phi \Phi \rangle ~~\textrm{propagators} &&\rightarrow~
[(n+p+3)- (k-m) -(2m-f+3))] D^2 \nonumber \\
\langle \bar{\Phi} \bar{\Phi} \rangle ~~\textrm{propagators}&&\rightarrow~
(k-m+f) \bar{D}^2 \nonumber
\eea
and counting the derivatives from the vertices, the total number of $D^2$'s
and $\bar{D}^2$'s is
\bea
&&D^2 : n-m+3p+2+f \nonumber \\
&&\bar{D}^2 : n-m+2p+2+f
\eea
In the most divergent configuration, the $D$-algebra produces a momentum
factor \begin{equation}
\Box^{n-m+p+f+1}
\end{equation}
and the condition
\begin{equation}
n \geqslant k+p \geqslant m+p
\end{equation}
The propagators give a factor
\begin{equation}
\Box^{-[2n-2m+2p+2f+3]}
\end{equation}
so that the corresponding momentum integral has dimension $8-2(n-m+f+2+p)$
and it is divergent if
\begin{equation}
n \leqslant 2 + m -f - p \leqslant 2 + m -p
\end{equation}
Since $n \geqslant m + p$ a divergence is present when $p \leqslant 1$.

%%%%%%%%%%%%%%%%%%%%%%%%%%%%%%%%%%%%%%%%%%%%%%%%%%%%%%%%
%%%%%%%%%%%%%%%%%%%%%%%%%%%%%%%%%%%%%%%%%%%%%%%%%%%%%%%
\vskip 18pt
\noindent
%---------- FIGURE TOP ------------
\begin{minipage}{\textwidth}
\begin{center}
\includegraphics[width=0.40\textwidth]{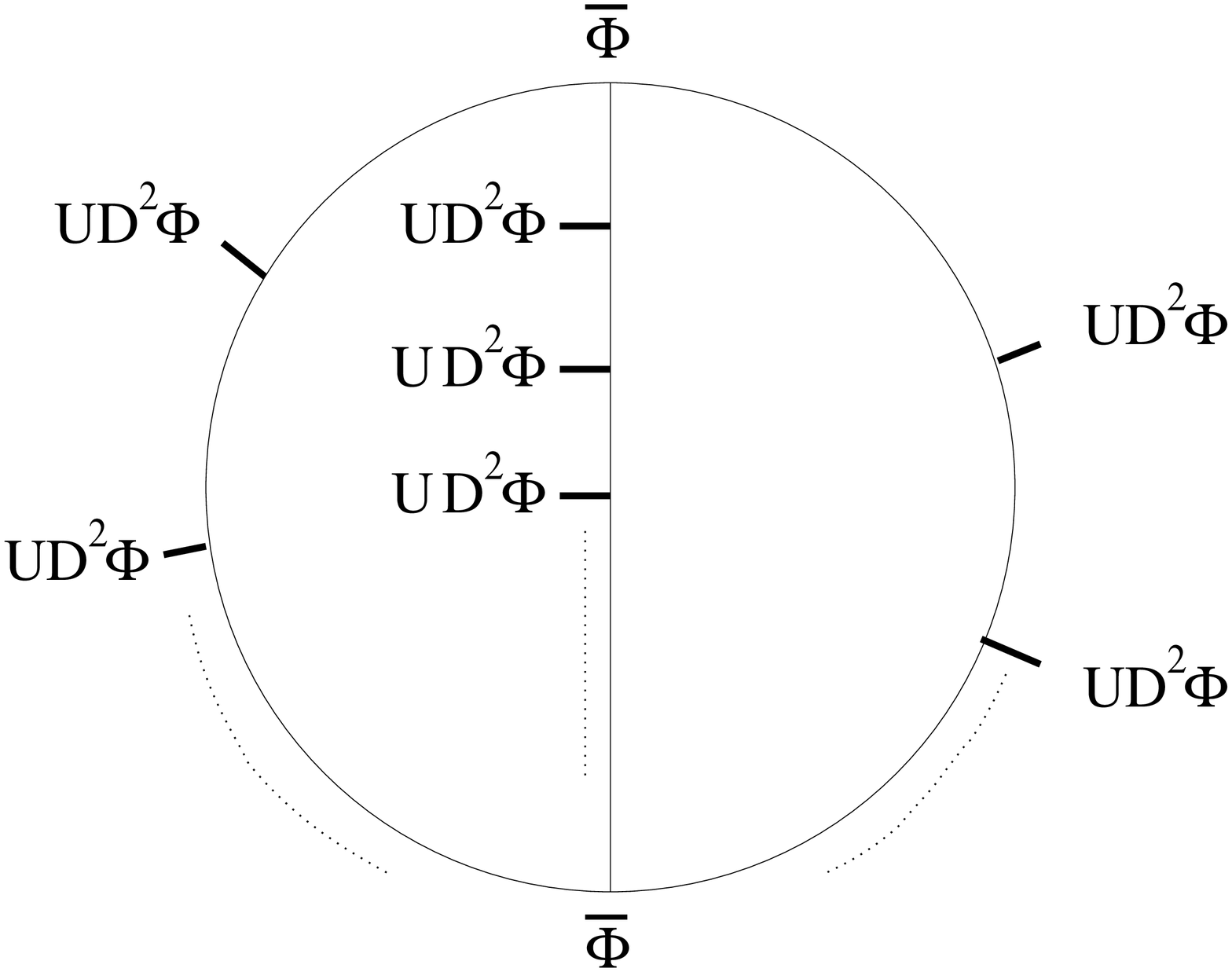}
\end{center}
\begin{center}
{\small{Figure B6}}
\end{center}
\end{minipage}
%---------- FIGURE END ------------
\vskip 20pt

B6) In this case we need introduce the parameter $f' = 0,\dots,6$ to count the
number of different configurations of external $\bar{\Phi}$ legs directly
connected to the internal $\bar{\Phi}^3$ vertices
and the configuration where the two internal $\bar{\Phi}^3$ vertices
are directly connected.

Since the number of propagators is
\bea
\langle \Phi \Phi \rangle ~~\textrm{propagators:}&&
[(n+p+3)- (k-m) -(2m-f'+6))] \nonumber \\
\langle \bar{\Phi} \bar{\Phi} \rangle ~~\textrm{propagators:}&&
(k-m+f')  \nonumber
\eea
the total number of covariant derivatives turns out to be
\bea
&&D^2 : n-m+3p+1+f' \nonumber \\
&&\bar{D}^2 : n-m+2p+f'
\eea
The most divergent configuration corresponds to the case where
a momentum factor
\begin{equation}
\Box^{n-m+p-1+f'}
\end{equation}
is produced by $D$-algebra, consistent with the condition
\begin{equation}
n \geqslant k+p+1 \geqslant m+p+1
\end{equation}
The propagators give a factor
\begin{equation}
\Box^{-[2n-2m+2p+2f']}
\end{equation}
and the momentum integral has dimension $8-2(n-m+f'+1+p)$.
It is divergent when
\begin{equation}
n \leqslant 3 + m -f' - p
\leqslant 3 + m -p
\end{equation}
Since $n \geqslant m + p +1 $, we obtain the condition $p \leqslant 1$.

In conclusion, B3 and B4 configurations never contribute to divergences, whereas the rest can produce divergent terms only when a single insertion of the cubic $U$ vertex is present.

As in the one-loop case, we can make an analogous analysis when
considering the insertions of $U (D^2 \Phi)^2$ vertices. Since the
$D$-algebra is identical we reach the conclusion that
at two loops only diagrams with one $U$-insertion (both quadratic or
cubic) can be divergent.
One can now proceed along the same lines to determine which are the actual
divergent diagrams for each topology. Precisely, for a given
configuration of internal and $U$ vertices, one can determine the number
and the distribution of external $\Phi$ and/or $\Phib$ legs associated
with divergent graphs. As a result one discovers that there cannot be
more that five external legs and the diagrams drawn in Figs. 5--17 exhaust
the complete set of two--loop divergent diagrams.

\end{document}